\documentclass[12pt]{article}
\pdfoutput=1
\usepackage{savesym}
\usepackage{cite}
\usepackage{wasysym}
\usepackage{amscd}
\usepackage{graphicx}
\usepackage{pifont}
\usepackage{float}
\usepackage{caption}
\usepackage{subcaption}
\usepackage[all]{xy}
\usepackage{multicol}
\usepackage{amsfonts}
\usepackage{picture}
\usepackage{amssymb}
\savesymbol{iint}
\savesymbol{iiint}
\usepackage{amsmath}
\usepackage{amsthm}
\restoresymbol{TXF}{iint}
\restoresymbol{TXF}{iiint}
\usepackage{color}
\usepackage{clay}
\usepackage{xcolor}
\definecolor{darkblue}{rgb}{0.0,0.0,0.3}
\usepackage[unicode=true,
bookmarks=true,bookmarksnumbered=false,bookmarksopen=false,
breaklinks=false,pdfborder={0 0 0},pdfborderstyle={},backref=false,colorlinks=true]
{hyperref}
\hypersetup{pdftitle={Calabi-Yau CFTs and Random Matrices},
	pdfauthor={},
	linkcolor=darkblue, urlcolor=darkblue, citecolor=darkblue, filecolor=black, linktoc=page}
\usepackage{setspace}
\usepackage[small]{titlesec}
\usepackage{microtype}
\usepackage{multirow}
\usepackage{wrapfig}
\usepackage{verbatim}
\usepackage{dsfont}
\usepackage[vcentermath]{youngtab}

\numberwithin{equation}{section}
\usepackage{float}
\restylefloat{table}
\allowdisplaybreaks
\usepackage{accents}

\usepackage[margin=1in]{geometry}

\PassOptionsToPackage{linecolor=blue,backgroundcolor=blue!25,bordercolor=blue,textsize=scriptsize}{todonotes}
\usepackage{todonotes}
  
\newcommand{\I}{{\mathrm i}}

\def\bar{\overline}

\def\1{{\mathds 1}}

\makeatletter
\g@addto@macro\bfseries{\boldmath}
\makeatother

\let\originalleft\left
\let\originalright\right
\renewcommand{\left}{\mathopen{}\mathclose\bgroup\originalleft}
\renewcommand{\right}{\aftergroup\egroup\originalright}

\makeatletter
\g@addto@macro\@floatboxreset\centering
\makeatother

\renewenvironment{thebibliography}[1]{%
\begin{oldthebibliography}{#1}%
\small%
\raggedright%
\setlength{\itemsep}{0pt}%
}%
{%
\end{oldthebibliography}%
}

\global\long\def\dd{\text{d}}%
\global\long\def\ii{\text{i}}%
\global\long\def\ee{\text{e}}%
\global\long\def\GL#1{\text{GL}(#1)}%
\global\long\def\vol{\operatorname{vol}}%
\global\long\def\Vol{\operatorname{Vol}}%
\global\long\def\Tr{\operatorname{Tr}}%
\global\long\def\op#1{\operatorname{#1}}%
\global\long\def\bC{\mathbb{C}}%
\global\long\def\bP{\mathbb{P}}%

\begin{document}

\begin{titlepage}

\begin{flushright}
\end{flushright}

\vfill

\begin{center}

{\setstretch{1.3}\Large\bf Calabi--Yau CFTs and Random Matrices
\par}

\vskip 1cm 

Nima Afkhami-Jeddi,\textsuperscript{a} Anthony Ashmore\textsuperscript{a,b} and Clay C\'ordova\textsuperscript{a}

\vskip 1cm

\textit{\small{}\textsuperscript{a}Enrico Fermi Institute \& Kadanoff Center for Theoretical Physics,\\
University of Chicago, Chicago, IL 60637, USA}  \\[.2cm]
\textit{\small\textsuperscript{b}Sorbonne Universit\'e, CNRS, LPTHE, F-75005 Paris, France}

\end{center}

\vfill

\begin{center}
\textbf{Abstract}
\end{center}

\begin{quote}
Using numerical methods for finding Ricci-flat metrics, we explore the spectrum of local operators in two-dimensional conformal field theories defined by sigma models on Calabi--Yau targets at large volume.  Focusing on the examples of K3 and the quintic, we show that the spectrum, averaged over a region in complex structure moduli space, possesses the same statistical properties as the Gaussian orthogonal ensemble of random matrix theory.
\end{quote}

\vfill 

{\begin{NoHyper}\let\thefootnote\relax\footnotetext{\tt \!\!\!\!\!\!\!\!\!nimaaj, ashmore, clayc@uchicago.edu}\end{NoHyper}}

\end{titlepage}

\microtypesetup{protrusion=false}
\tableofcontents
\microtypesetup{protrusion=true}

\section{Introduction and Summary}

\begin{figure}
	\includegraphics{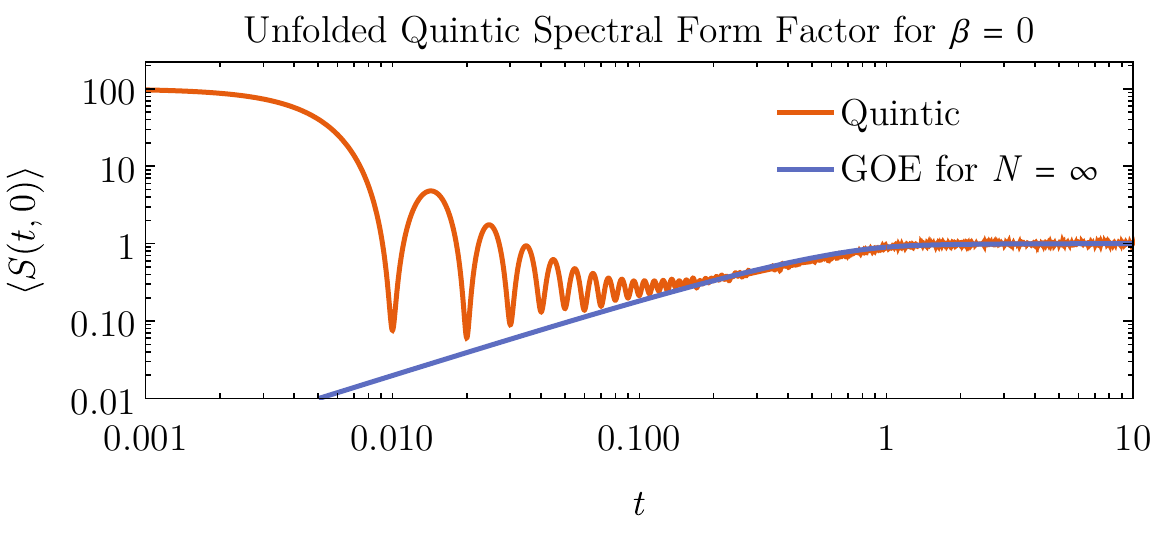}\\
	\includegraphics{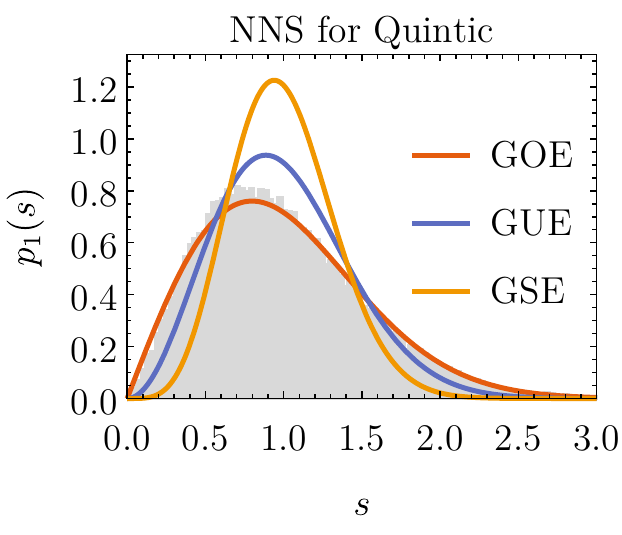}\qquad\includegraphics{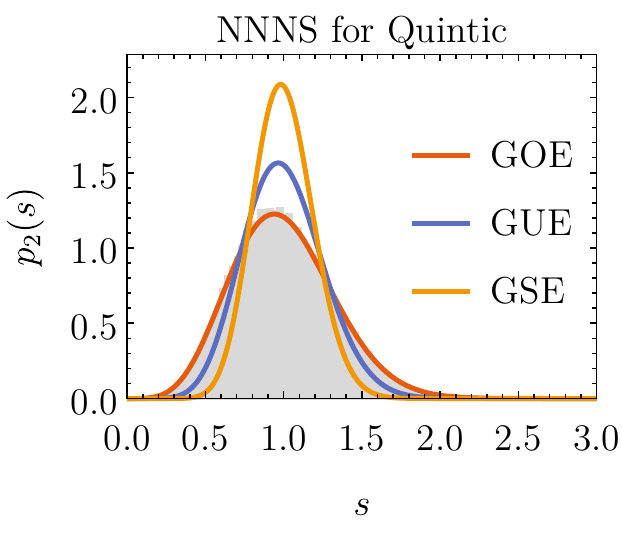}\\
	\includegraphics{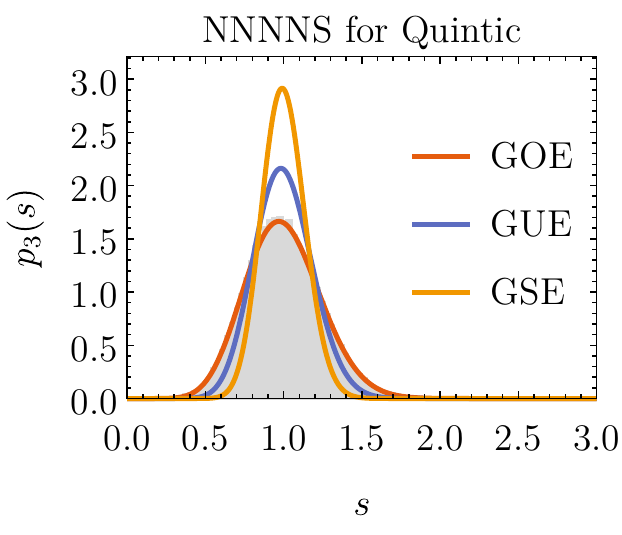}\qquad\includegraphics{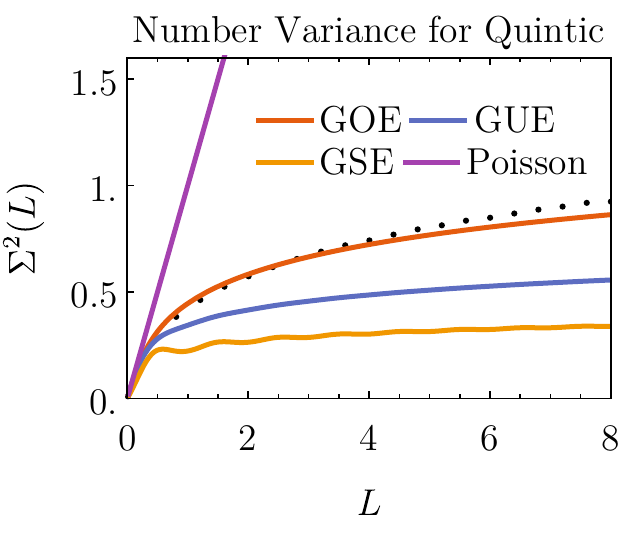}
	\caption{The top row shows the unfolded spectral form factor for $\beta=0$ for quintic Calabi--Yau threefolds. The orange curve shows an ensemble average consisting of 1,000 samples in complex structure moduli space with each sample containing the lowest-lying 100 eigenvalues. The dip, ramp and plateau are all present. The second and third rows display certain spectral statistics together with fits to GOE.  (Other matrix ensembles are also shown for comparison.) The second row shows the nearest-neighbor level spacings (NNS) and the next-to-nearest (NNNS). The third row shows the next-to-next-to-nearest (NNNNS) level spacings and the number variance.}
	\label{fig:QF}
\end{figure}

In this paper, we use numerical methods to investigate two-dimensional conformal field theories defined by sigma models on Calabi--Yau targets. Specifically, we compute the low-lying spectrum of the Laplacian on Calabi--Yau manifolds equipped with Ricci-flat metrics. In the large-volume limit, the corresponding excitations come from string momentum modes, and so this calculation gives the spectrum of local operators in the worldsheet CFT. Computing this spectrum for many different choices of moduli gives an ensemble of CFT data that we can analyze statistically. In particular, we show that the spectrum averaged over a region in complex structure moduli space enjoys the same statistical properties as the Gaussian orthogonal ensemble (GOE) of random matrix theory (RMT).  An example of our results for the quintic Calabi--Yau threefold is shown in Figure \ref{fig:QF}.

\subsection{Random Matrices and Quantum Chaos}

Random matrices are a hallmark of quantum chaos (see \cite{mehta2004random,Br_zin_1997,RevModPhys.53.385,Guhr:1997ve} and references therein for a review). The seminal conjectures of Bohigas, Giannoni and Schmit~\cite{B1,B2} codified this by postulating that systems whose classical limits show sufficient ergodicity display random matrix statistics in their quantum energy levels.  Some of the most prominent features of quantum chaos are the repulsion of nearby energy levels and the long-range rigidity of the spectrum, both of which contrast sharply with the statistics of a Poisson random process. (See the nearest-neighbor level spacings (NNS) and $\Sigma^{2}$ plots in Figure \ref{fig:QF}.)

The study of quantum chaos goes back more than four decades and has a rich interplay with classical chaotic dynamics, semiclassical physics, asymptotic methods, random matrix theory and spectral theory. Let us recall some of the key ideas.

In classical mechanics, integrability and chaos are well-defined notions. In particular, in the absence of sufficiently many conserved quantities (via symmetries), one generically expects the strongest form of chaos, where the classical dynamics is deterministic but the system is so unstable that small changes in initial conditions lead to large variations in its long-time evolution. This is often called the ``butterfly effect'' and can be seen, for example, in Bunimovich's billiards~\cite{cmp/1103904878}.

This form of chaos does not apply to quantum systems, which do not have a well-defined notion of phase space trajectories. In quantum mechanical systems, instead of position and velocity, one discusses states and their energy spectra. A naive definition of quantum chaos, i.e.~whether two ``close'' initial states $\psi_1(\vec{r},t=0)$ and $\psi_2(\vec{r},t=0)$ diverge exponentially from one another with time, is not useful in quantum mechanics, essentially due to linearity of the Schr\"odinger equation (the overlap of two states evolved with the same Hamiltonian is constant in time).\footnote{The most natural generalization of the butterfly effect to quantum systems is found by examining out-of-time order correlators.} Instead, a key question of quantum chaos is whether a quantum system displays qualitatively different behavior if its classical analogue is chaotic.

There are now many examples of how classical chaos affects the quantum dynamics of a system. In particular, this has been formalized via semiclassical trace formulae, such as the Gutzwiller trace formula for systems in a chaotic regime~\cite{doi:10.1063/1.1665596}. Roughly speaking, the essential idea is that the density of states of the quantum system can be written as a sum of two contributions: a slowly varying term associated with the classical energy of the system and an oscillating term which captures quantum fluctuations. The Gutzwiller trace formula then connects the quantum contribution to a classical object, which is computed from periodic orbits of the classical system. Thus, this gives a link between a quantum system and its classical analogue by expressing its energy spectrum in terms of periodic orbits (with longer orbits needed to resolve smaller energy differences). 

The existence of these trace formulae imply that classical chaotic dynamics should affect the quantum mechanical properties of a system, at least in some qualitative way. In `77, Berry and Tabor conjectured that generic integrable systems have energy levels that follow Poisson statistics~\cite{10.2307/79349}. In `84, Bohigas, Giannoni and Schmit~\cite{B1,B2} extended this to the chaotic regime, conjecturing that the energy spectrum of a classically chaotic system should display the eigenvalue repulsion and spectral rigidity characteristic of RMT.
In other words, RMT should give a statistical description of the spectrum and eigenfunctions of the corresponding quantum Hamiltonian. This conjecture has been verified in a variety of systems, with only non-generic systems known to violate it~\cite{PhysRevLett.69.2188}. Thus, RMT-like statistics for the energy spectrum is often taken to be a defining property of quantum chaos and a signature of the kind of classical dynamics underlying the quantum system.\footnote{Indeed, a more precise definition of quantum chaos still appears to be missing~\cite{whatischaos,zelditchchaos}.}

Despite this long history, we note that there are few universal or rigorous results on chaos and ergodicity, other than for 2d billiards~\cite{Katok}, compact manifolds with negative curvature~\cite{Hedlund,Hopf1,Hopf2,Asnov}, and related systems. Indeed, even for the simple example of billiard motion in a regular tetrahedron, it is not known whether the trajectories are ergodic~\cite[\S1.9.3]{Berger}. Furthermore, it is not known whether ergodicity is sufficient for random matrix statistics in the resulting spectrum. For example, it is known that generic Riemann surfaces have ergodic geodesics~\cite{Asnov}, however it is not known how to show that this implies their spectra will display GOE statistics (see, for example, \cite[\S9.13.2]{Berger}).

Since chaos is generic in physical systems, the general features of random matrix theory are ubiquitous and have found applications in nuclear physics~\cite{10.2307/1970079,PhysRev.120.1698,PhysRev.123.1293,Haq19821086}, billiards~\cite{B1,Gutzwiller_1990,BALAZS1986109}, and the quantum hall effect~\cite{PhysRevB.43.8641} to name but a few. Recently, a particular diagnostic, the spectral form factor (SFF), has been used to investigate spectral correlations in quantum many-body systems such as the SYK model~\cite{Sachdev:1992fk,Kitaev1,Kitaev2,Kitaev3} as well as in the context of black hole physics and gravitational systems~\cite{Saad:2018bqo,Garcia-Garcia:2016mno,You_2017,Cotler:2016fpe,Gharibyan:2018jrp}.  In this situation, the essential result is that the black hole energy levels are discrete, non-degenerate and chaotic.  At late times, this implies that correlations functions cannot decrease, but instead must oscillate \cite{Maldacena:2001kr,Dyson:2002nt,Barbon:2003aq,Papadodimas:2015xma}.  This behavior, suitably averaged, is precisely captured by random matrix statistics and leads to features known colloquially as the \emph{ramp} and the \emph{plateau}.  (See the SFF plot in Figure \ref{fig:QF}.)

\subsection{Conformal Field Theories and Calabi--Yau Metrics}

The appearance of random matrix theory in the setting of quantum gravity suggests, via holography, that generic conformal field theories might also display such features in their spectrum.  However, to date, few explicit field theory calculations have been carried out.  Notable exceptions are the averages of 2d free field theories discussed in \cite{Afkhami-Jeddi:2020ezh,Maloney:2020nni,Benjamin:2021wzr} where exact calculations give rise to a Chern--Simons-like theory of gravity.  By contrast, in this work, we analyze genuine interacting theories where RMT statistics are expected.   

Our focus is (super)conformal field theories defined by sigma models into Calabi--Yau targets; explicitly, we consider the K3 surface and the quintic threefold.  These theories have been well studied from a variety of points of view as they give rise to rich target spaces for string theory.  Most calculations in these theories involve observables protected by supersymmetry which can often be computed even at strong coupling.  (See e.g.~\cite{Witten:1982df,Eguchi:1987wf,Eguchi:1988vra,Cecotti:1992qh,Kawai:1993jk,Dijkgraaf:1996it,Benini:2013nda,Lin:2015dsa}.) There has also been work exploring these conformal field theories making use of the modular properties of the partition function and from the perspective of the conformal bootstrap~\cite{Hellerman:2009bu,Keller:2012mr,Friedan:2013cba,Lin:2015wcg,Collier:2016cls,Lin:2016gcl}.

In each of these previous analyses, the focus has been on the properties of a few low-dimension operators (such as those in the chiral ring).  By contrast, to compare with the predictions of random matrix theory, one needs access to hundreds of (non-BPS) eigenstates so that their statistics can be cleanly diagnosed.  To achieve this precision, we will work in the large-volume regime of the sigma model. In this weakly coupled limit, the theory reduces to the quantum mechanics of a particle moving in the Calabi--Yau target. We can then determine the spectrum by solving for the eigenvalues of the Laplace operator on this manifold. From this perspective, the problem we are looking at is not completely new, but is instead an advanced, higher-dimensional cousin of those previously studied in quantum chaos and discussed above.

A fundamental challenge in carrying out this calculation is the fact that Ricci-flat metrics on Calabi--Yau manifolds (required to ensure a vanishing beta function in the conformal field theory) are generally not known analytically.\footnote{See \cite{1810.10540, 2006.02435, 2010.12581} for recent progress on computing K3 metrics analytically.}  As a result, we proceed numerically.  The past decade has seen a number of techniques introduced for computing numerical K\"ahler--Einstein metrics, with a particular focus on Ricci-flat examples on Calabi--Yau manifolds. These include a lattice approach~\cite{hep-th/0506129}, spectral methods~\cite{hep-th/0612075,0712.3563,math/0512625,0908.2635}, and most recently neural networks~\cite{2012.04656,2012.04797,2012.15821}. In this paper, we will use the spectral method which has become known as ``Donaldson's algorithm''~\cite{math/0512625} to compute the Ricci-flat metrics numerically. We then compute the scalar Laplacian using the method laid out in \cite{0805.3689,Ashmore:2020ujw}, solving for the spectrum of operators for many choices of complex structure moduli, and then analyze the resulting distribution of eigenvalues.  

As expected for a complex system, the spectrum at a fixed value of couplings, in this case complex structure moduli, is highly erratic.  To detect structure, we pass to an ensemble by sampling over a region in complex structure moduli space.\footnote{Our averaging procedure is rather rudimentary, weighting each point equally.  One could also consider averaging using the Weil--Petersson metric on the moduli space, as was done for the torus in \cite{Afkhami-Jeddi:2020ezh,Maloney:2020nni,Benjamin:2021wzr}.} After unfolding, the resulting ensembles in the examples we consider show good agreement with the expectations of the large-$N$ GOE ensemble to within the numerical precision available in our calculations.  

\subsection{Comments and Future Directions}

Since the generic behavior of physical systems is thought to be chaotic, it is perhaps not surprising that one finds RMT statistics in sigma models. For example, taking appropriate random metrics on the target space, one would expect the spectrum of the Laplacian to obey RMT. What \emph{is} surprising is that one sees this chaotic behavior even for the restricted class of Ricci-flat (or K\"ahler--Einstein) metrics which are relevant for 2d CFTs.\footnote{Note that repeating the numerics for a torus sampled over its complex structure modulus does \emph{not} lead to RMT statistics. Instead, one finds Poisson statistics, in agreement with the general results of \cite{10.2307/79349,10.1007/978-3-0348-8266-8_36} for a system that is classically integrable.} Comparing to the conjectures of \cite{B1, B2}, the presence of GOE statistics strongly suggests that classical geodesic motion on Ricci-flat Calabi--Yau manifolds is ergodic. This agrees with the general expectation that particle motion on manifolds without isometries is chaotic. We expect that our results will hold for any $d>2$ Ricci-flat Calabi--Yau geometries that come in families labeled by continuous moduli, giving a huge number of 2d SCFTs whose low-lying spectra should display chaos. It would be interesting to explore this connection further and see if it hints at whether or not there is a useful physical interpretation for ensemble-averaged CFTs. 

It is possible to extend our work to the full $(p,q)$-form spectrum, which would correspond to including operators whose scaling dimensions are shifted upwards by $(p+q)/2$. As we discuss in Section \ref{sigmasec}, in the large-volume limit, these operators are well separated from the scalar spectrum that we examine, and so it is consistent to truncate them. If one wants to relax the large-volume limit, it would be necessary to include these modes.

It would also be interesting to examine the distribution of Yukawa-like couplings over moduli space.\footnote{One could also then compare these with the general bounds for Kaluza--Klein modes derived in \cite{Hinterbichler:2013kwa}.} In principle, this data is computable from overlap integrals of the Laplacian eigenmodes (coupled to a gauge bundle). In practice, however, the evaluation and storage of a large enough number of these with which to do statistics is computationally intensive. Physically, such data would be extremely useful, especially in light of the fact that fixing moduli has proven to be so difficult, as one could then ask questions about what kind of four-dimensional physics is possible for a given Calabi--Yau, allowing a true exploration of the landscape. Furthermore, while analytically computing non-BPS quantities, such as normalized Yukawa couplings or K\"ahler potentials, is currently out of reach, the ensemble-averaged versions may have simpler mathematical descriptions. We hope to return to this in the future. We also note that the idea of looking at the statistics of string vacua is an old one, whether via distributions of flux vacua~\cite{Ashok:2003gk,Douglas:2004zu,Douglas:2003um,Denef:2004ze,Denef:2004cf,Distler:2005hi,Podolsky:2008du} or ``random'' scalar potentials~\cite{Marsh:2011aa,Eckerle:2016hzt}. The novelty of our approach is the access to the non-BPS data of the string/CFT spectrum. 

The Yukawa-like couplings mentioned above are a special case of OPE coefficients in the CFT. Associativity of the operator algebra imposes constraints on the OPE coefficients of conformal field theories. The presence of the vacuum in the spectrum of a CFT, together with these constraints, results in universal behavior for the averaged OPE coefficients\cite{Collier:2019weq}, just as it implies universal behavior for the asymptotic averaged density of states, given by Cardy's formula~\cite{Cardy:1986ie}. This is consistent with the expectation that chaotic CFTs exhibit  dynamics that are compatible with the eigenstate thermalization hypothesis. However, the universal behavior of the OPE coefficients, like Cardy's formula, is a good description only if the scaling dimensions of the operators under consideration are sufficiently large. Since our methods enable us only to access the OPE coefficients corresponding to the low-lying operators, we do not expect to observe universality in the OPE data. 

Note that here we observe RMT behavior in the spectral statistics of the low-lying CFT operators which survive the large-volume limit. This is in contrast to recent works~\cite{1611.04592,Cotler:2016fpe,1706.05400,Gharibyan:2018jrp} where RMT behavior is expected to arise from the chaotic dynamics of black holes, and is therefore observed in the spectrum of the heavy microstates with a dense spectrum accounting for the black hole entropy. In cases where a holographic dual exists, it is expected that observables such as the SFF receive contributions from saddles corresponding to wormholes in the gravitational path integral, providing a dual gravitational description for various RMT features~\cite{Cotler:2020ugk,Saad:2018bqo,Saad:2019lba,Collier:2021rsn}. It is not clear whether the light sector of the ensemble-averaged sigma model admits such a description. It would be interesting to explore whether the ensemble average over the full sigma model admits a dual description.\footnote{See \cite{Benjamin:2021wzr} for progress made in this direction.}

It is also worth mentioning that there is a slight tension with CFT consistency conditions if the low-lying operators are strictly described by RMT statistics. This is because according to the RMT description, there is a finite but small probability that the lightest operator in the theory is arbitrarily heavy which conflicts with modular invariance of the torus partition function~\cite{Hellerman:2009bu,Collier:2016cls}. Nevertheless, as we will show, the light operators of the sigma model clearly exhibit statistics indicative of chaotic behavior.

We start with review of RMT and some spectral diagnostics in Section \ref{rmtsec}. We then outline the connection between sigma models, CFTs and the spectrum of the Laplacian in the large-volume limit in Section \ref{sigmasec}, and give an overview of our numerical method for numerically computing Ricci-flat metrics and the spectrum of the Laplacian in Section \ref{laplacian}. Finally, in Section \ref{eq:results} we present our results for the spectra of genus-3 curves, K3 surfaces and quintic threefolds, finding agreement with GOE statistics in all cases.

\section{Spectral Statistics and Random Matrix Theory}
\label{rmtsec}

In this section, we begin by recalling the concept of ensemble averages of physical systems, followed by reviewing properties of the Gaussian orthogonal ensemble (GOE) of random matrices.  This is the appropriate ensemble for a real symmetric matrix corresponding to a Hamiltonian with time-reversal symmetry. It is expected that eigenvalue statistics of quantum chaotic systems with this symmetry are well described by the GOE. In the figures that follow for comparison we sometimes plot various quantities for the Gaussian unitary ensemble (GUE) and Gaussian symplectic ensemble (GSE), though we do not present a detailed discussion of these models.\footnote{The interested reader can find an even larger classification of symmetries of random matrices in terms of Riemannian symmetric spaces in \cite{Zirnbauer10}.} We then discuss various statistics that capture short- or long-range correlations in spectra, and give the corresponding distributions for GOE in the large-$N$ limit. These statistics will be used in Section \ref{eq:results} to analyze the scaling dimensions of scalar operators in certain CFTs, so it is useful to introduce them first within the well-known context of RMT.

Consider a single instance of a quantum system with energy levels $E_{n}=(E_{1},E_{2},\ldots)$. These might also be the eigenvalues of a single matrix from an ensemble. The density of states is defined to be 
\begin{equation}\label{eq:density_E}
	\rho(E)=\sum_{n}\delta(E-E_{n})~.
\end{equation}
When we have an ensemble of systems with a corresponding set of Hamiltonians and their spectra, such as an ensemble of random matrices, we can define ensemble-averaged quantities to be simply the average over the ensemble. Unless otherwise specified, we use angle brackets to denote such an ensemble average: for example, $\langle\rho(E)\rangle$ is the averaged density of states.

Note that if one has only a single instance of a system, one can still define an average by integrating over an energy window:
\begin{equation}
	\overline{\rho(E)}=\frac{1}{\delta}\int_{E-\delta/2}^{E+\delta/2}\dd E'\,\rho(E')~,
\end{equation}
where $\delta$ should be large enough to smooth out fluctuations but small enough to be negligible on the scale at which the average itself varies (so that the result is somewhat independent of $\delta$). Since the systems we consider naturally come from ensembles, we will not use this alternative formulation.\footnote{As mentioned in \cite{chao-dyn/9606010}, energy averaging is appropriate when the results depend weakly on the averaging window $\delta$. This is usually the case if: a) there are a large number of energy levels in the average; b) $\delta$ is classically small. Due to numerical limitations, we are able to compute only the lowest-lying 100 or so eigenvalues in our examples. Thus, if we try to include a ``large'' number of levels in the average, the averaging window is necessarily not small compared with the range of the eigenvalues. Because of this, the dependence on $\delta$ is not weak, and hence ensemble averaging is more appropriate.}

The appearance of random matrices in the study of chaotic systems is well established, though perhaps surprising at first sight. The first observation is that for random matrix ensembles, in the large matrix limit (large $N$), spectral averages over a single matrix are equivalent to averages over the full ensemble. This allows one to search for RMT-like behavior in systems by considering ensembles of theories and then averaging. Work in recent decades has suggested that RMT should be thought of as describing somewhat generic properties of these systems. One should then ask whether a particular instance of the system belongs to a subset of the ensemble that has non-generic spectral statistics. Chaotic systems should then be thought of as systems where we lack a priori information and thus have no reason to think of them as being non-generic. Conversely, systems which are integrable classically should be thought of as non-generic. RMT is thus often a ``null hypothesis''. In a sense, a main result of this paper is that the low-lying spectrum of certain CFTs in the large-volume limit is sufficiently generic to display RMT behavior.

There are several diagnostics of spectral correlations focusing on different features of the spectrum which have proven useful in verifying RMT predictions. We focus on aspects relevant to our analysis of the spectrum of Laplace operator.  We will provide only a brief overview of the relevant concepts. See \cite{mehta2004random,Guhr:1997ve} for a more comprehensive discussion.

For an $N\times N$ random matrix drawn from the GOE, the joint probability density for the eigenvalues $E_{i}$ is given by
\begin{equation}\label{jpd}
P_N(E_1,...,E_N)=C_{N}\exp\left(-\frac{1}{2}\sum_{i=1}^N E_i^2\right)\left(\prod_{1\leq j<k\leq N}|E_j-E_k|\right)~,
\end{equation}
where $C_{N}$ is a normalization constant.  This distribution exhibits the key features of Gaussian falloff of the individual $E_{i}$, as well as the Vandermonde repulsion of pairs of eigenvalues.  The $k$-point correlation functions which encode partial probability distributions where only $k$ eigenvalues are observed can then be expressed as
\begin{equation}
R_k(E_1,...,E_k) \propto \int \left(\prod_{i=k+1}^N \dd E_i \right)P_N(E_1,...,E_N)~.
\end{equation} 

Analytic expressions for the correlation functions may be obtained in the $N\rightarrow \infty$ limit. Famously, the Wigner semi-circle distribution describes the density of states in this limit:
\begin{equation}\label{density}
\rho(E)\equiv  N R_1(E)=\frac{1}{\pi}\sqrt{2N-E^2}\hspace{.5in}\text{for }N\rightarrow\infty~.
\end{equation}
Below, we focus on the large-$N$ limit since we are interested in comparing the spectral properties of a Hamiltonian in an infinite-dimensional Hilbert space, albeit with a practically limited cut-off, to RMT predictions. As we mentioned earlier, the large-$N$ results also capture the leading behavior of ensemble-averaged finite-$N$ matrices.

In general applications of RMT to chaotic Hamiltonians, the density of states \eqref{density} is not a realistic model.  Instead, the spectral correlations are captured by the statistics of the distribution \eqref{jpd}. Given a physical system, the eigenvalue correlations will have a system-dependent contribution and a random universal part. Generally, it is only the latter part that can be used for comparison between different systems or different matrix ensembles. To isolate this universal part, we remove the dependence on the mean density of states by ``unfolding'' the spectrum, giving a new sequence of eigenvalues with the same fluctuation properties but with a mean density equal to one~\cite{Guhr:1997ve}. This is done by separating the eigenvalues by an amount inversely proportional to the mean density.  Specifically, we introduce a variable
\begin{equation}\label{eq:unfold}
\alpha(E)\equiv \int_{-\infty}^{E}\dd E'\,R_{1}(E')~,
\end{equation} 
which counts the number of eigenvalues with value less than $E$.  We then transform all correlators to functions of the $\alpha$ variables by introducing unfolded correlation functions $G_{k}(\alpha_{1}, \alpha_{2}, \cdots, \alpha_{k})$ obeying
\begin{equation}
G_{k}\bigl(\alpha_{1}(E_{1}), \alpha_{2}(E_{2}), \cdots, \alpha_{k}(E_{k})\bigr)\equiv \frac{R_{k}(E_{1}, E_{2}, \cdots, E_{k})}{R(E_{1})R(E_{2})\cdots R(E_{k})}~.
\end{equation}
In particular, the unfolded one-point function $G_{1}(\alpha)$ is constant.  The unfolded correlation functions can thus be interpreted as conditional probability distributions. In what follows, we primarily consider the unfolded spectrum and correlators in the large-$N$ limit.

The discussion above assumes a single instance of an $N\times N$ matrix, however one can also unfold the spectra of an ensemble of random matrices. To do this, we simply define $\alpha(E)$ to be the ensemble average of the number of eigenvalues less than $E$:
\begin{equation}
	\alpha(E) \equiv \left\langle\int_{-\infty}^{E}\dd E'\,\rho(E')\right\rangle~,
\end{equation}
that is, one unfolds the spectrum using the average eigenvalue density $\langle \rho(E) \rangle$.

\subsection{Short-Range Correlations}

The short-range correlations in the spectrum may be probed by studying the distribution of nearest-neighbor spacings (NNS) between the eigenvalues, which we denote by $p_1(s)$. The NNS describes the probability that two randomly chosen consecutive eigenvalues in the spectrum have a separation $s$. The NNS distribution may be defined using the joint probability distribution, and in general depends on all $(k\geq 2)$-point correlations in the spectrum. Exact analytic results can be expressed in terms of fast-converging infinite products in the large-$N$ limit for the Gaussian ensembles\cite{mehta2004random}. However, for our purposes it suffices to use Wigner's approximation to the spacing distribution given by
\begin{equation}
p_{1}(s)=\frac{\pi}{2}s \exp\left(-\frac{\pi}{4}s^2\right)~.
\end{equation}
The fact that the maximum of this function occurs away from $s=0$ reflects eigenvalue repulsion, which is expected due to the Vandermonde determinant appearing in the probability density for the eigenvalues. This should be contrasted with the nearest-neighbor statistics arising from an uncorrelated Poisson distribution, which is described by a simple exponential decay $p(s)\sim \ee^{-s}$.

More generally, we may also be interested in  higher-order level spacings, i.e.~next$^{k-1}$ nearest-neighbor spacings. A Wigner-like ansatz determines these distributions $p_{k}(s)$ to be~\cite{mehta2004random}
\begin{equation}
p_{k}(s)= s^{a_{k}} \exp\left(-b_{k}s^2\right)~,
\end{equation}
where the omitted constant of proportionality is fixed by demanding $p_k(s)$ integrates to one, and the $k$-dependent parameters are given by
\begin{equation}
a_{k}=\frac{k^{2}+3k-2}{2}~, \hspace{.5in}b_{k}=\Gamma^{2}\left(\frac{(2+k)(1+k)}{4}\right)\left/\Gamma^{2}\left(\frac{(3+k)k}{4}\right)\right.~.
\end{equation}

\subsection{Long-Range Correlations}

There are various diagnostics which probe the long-range correlations in the spectrum.  Here, we focus on quantities derived from the two-point correlation function $G_{2}(r)$ with $r=\alpha_{1}-\alpha_{2}\geq 0$.   At large $N$, away from the edges of the spectrum, one has  \cite{mehta2004random}  
\begin{equation}\label{twopt}
G_{2}(r)=1-\frac{\sin^{2}(\pi r)}{\pi^{2}r^{2}}-\frac{\dd}{\dd r}\left(\frac{\sin(\pi r)}{\pi r}\right)\int_{r}^{\infty}\dd y\, \frac{\sin(\pi y)}{\pi y}~.
\end{equation}
The constant $1$ above can be viewed as the contribution of the disconnected product of one-point functions.  The connected correlator decays as a power law ($\sim r^{-2}$) illustrating the long-range correlations.  In practice, since we deal with finite-size systems, we will need to be aware of finite-$N$ modifications to the expression above. There are a number of statistical quantities derived from the two-point function, two of which we now review.

\subsubsection{Number Variance}

 Let $\eta(L,\alpha)$ be the number of levels in the interval $[\alpha,\alpha+L]$.  The number variance $\Sigma^{2}(L)$ is then
\begin{equation}
\Sigma^{2}(L)=\overline{ \eta(L,\alpha)^{2}} -\overline{ \eta(L,\alpha)}^{2}~,
\end{equation}
where here the bars indicate an average over the starting point $\alpha$.  By definition, on the unfolded scale $\overline{ \eta(L,\alpha)}=L$.  Thus, in an interval of length $L$, we expect on average $L\pm \sqrt{\Sigma^{2}(L)}$ levels.  It is straightforward to derive a relationship between $\Sigma^{2}$ and the (connected) two-point function:
\begin{equation}
\Sigma^{2}(L)=L+2\int_{0}^{L}\dd r\, (L-r)(G_{2}(r)-1)~.
\end{equation}
In particular, using expression \eqref{twopt} for the correlator we see that for large $L$ and $N$ we have in the GOE
\begin{equation}
\Sigma^{2}(L)\approx \frac{2}{\pi^{2}}\log(L)+O(1)~.
\end{equation}
This logarithmic growth is an indication of spectral rigidity and should be contrasted with a Poisson process where the connected correlator vanishes, and $\Sigma^2$ grows linearly with $L$. The expression for $\Sigma^2(L)$ including subleading corrections in $L$ can be found in, for example, \cite[Eq.~5.13]{RevModPhys.53.385}. We use the full expression when checking our numerical results in Section \ref{eq:results}.

\subsubsection{Spectral Form Factor}

The spectral form factor (SFF), $S(t,\beta),$ is another useful diagnostic for probing correlations present in the spectrum~\cite{Haake10}. Given a Hamiltonian $H$ for some system, one can analytically continue the thermal partition function $Z(\beta)$ to a function of real time $t$ and inverse temperature $\beta$:
\begin{equation}
	Z(t,\beta) = \Tr \ee^{-(\beta+2\pi\ii t)H}~.
\end{equation}
The time average of this function vanishes, with $Z(t,\beta)$ itself oscillating wildly at late times (compared with the spacing of energy levels). The size of the fluctuations are captured by the SFF:
\begin{equation}\label{eq:SFF_def}
S(t, \beta)\equiv \frac{\left|Z(t,\beta)\right|^2}{Z(0,2\beta)}=\frac{\left|\Tr \ee^{-(\beta+2\pi \I t)H}\right|^2}{\Tr \ee^{-2\beta H}}~.
\end{equation}
Following \cite{Cotler:2016fpe}, we will take the ensemble average of the SFF, $\left\langle S(t,\beta) \right\rangle$, to be the annealed quantity, corresponding to averaging the numerator and denominator separately.

As usual, one can expand the partition function as a sum over energy eigenstates leading to
\begin{equation}\label{eq:Z_int}
\left|Z(t,\beta)\right|^2=\int \dd E_{1} \dd E_{2}\, \rho(E_{1})\rho(E_{2})\ee^{-\beta(E_{1}+E_{2})+ 2\pi \I t(E_{1}-E_{2})}~,
\end{equation}
where the density of eigenstates of the Hamiltonian $H$ is given by \eqref{eq:density_E}. Looking at this expression, we see that the SFF contains information about correlations between pairs of eigenvalues that may be close together or well separated, thus giving information about both short- and long-range correlations. In a chaotic system, the energy differences are mutually irrational, leading to a highly oscillatory function for large values of $t$.  Averaging over these fluctuations, the late-time behavior of $\left|Z(t,\beta)\right|^2$ is dominated by terms where $E_{n}\approx E_{m}$.  Thus:
\begin{equation}\label{lims}
S(t, \beta)\xrightarrow{\text{avg over }t\to\infty}\frac{Z(0,2\beta)}{Z(0,2\beta)}=1~,
\end{equation}
where the choice of denominator in \eqref{eq:SFF_def} ensures this averages to $1$.

Given an ensemble of theories with corresponding Hamiltonians and spectra of energies, we can pass to an ensemble average by expressing the expectation value of the product of densities in terms of the connected correlator.  On the unfolded energy scale this leads to 
\begin{equation}
\bigl\langle  \rho(\alpha_{n})\rho(\alpha_{m}) \bigr\rangle = \delta(\alpha_{n}-\alpha_{m})+\bigl(G_{2}(\alpha_{n}-\alpha_{m})-1\bigr)~.
\end{equation}
Thus the ensemble-averaged partition function can be expressed in terms of the (analytically continued) Fourier transform of the connected two-point correlator:
\begin{equation}\label{SFFRMT}
\left\langle \left|Z( t, \beta)\right|^2\right\rangle=\frac{1}{2\beta}+\Re\left[\frac{1}{\beta}\int_{0}^{\infty} \dd r\, \bigl(G_2(r)-1\bigr)\ee^{-\beta r+2\pi \I r t} \right]~.
\end{equation}
The resulting averaged spectral form factor for chaotic systems has several distinct features of interest known as the ``dip'', ``ramp'' and ``plateau''. These features are clearly visible in Figure \ref{fig:sffnvgoe}, which shows the averaged SFF for the case of a GOE.  Each of these features probes different aspects of the spectrum:
\begin{figure}
	\includegraphics{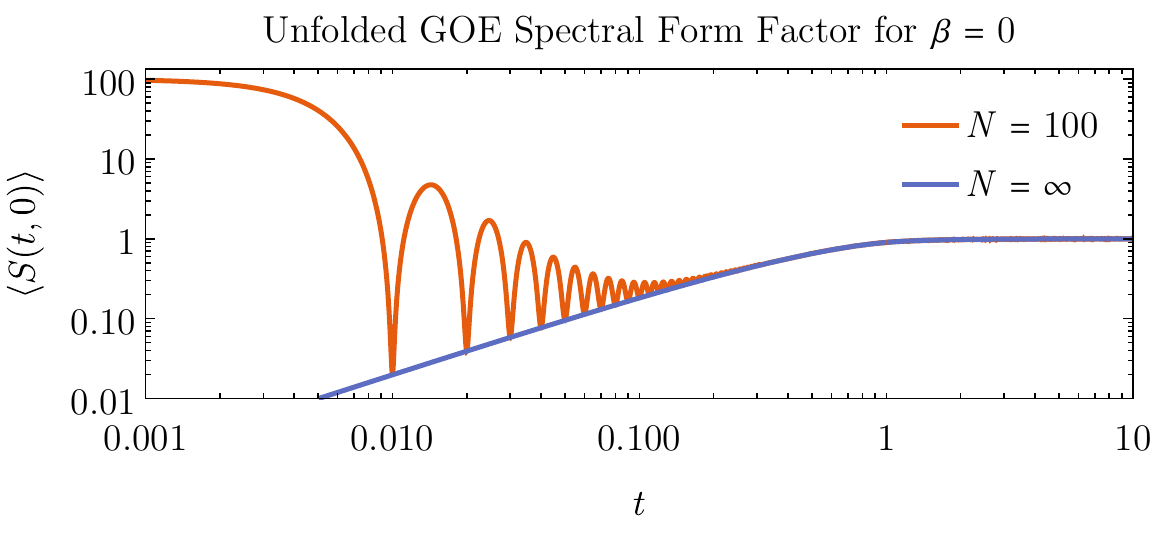}
	\caption{Unfolded finite-$N$ vs infinite-$N$ GOE spectral form factors for $\beta=0$.  The orange curve shows an ensemble average consisting of the eigenvalues of 100,000, $100\times 100$ random symmetric matrices.  After the dip, there is a period of approximately linear growth moderated by oscillations of decreasing amplitude, followed by a plateau. The $\beta\to0$ limit of equation \eqref{eq:Z_infinite_N} exactly matches the orange curve, so we have not plotted it separately. The blue curve shows the spectral form factor for $N\to\infty$ and $\beta\to0$ as given by \eqref{eq:Z_infinite_N}.  Only the ramp and plateau are present. The effect of non-zero $\beta$ is to damp the oscillations on the ramp region. }
	\label{fig:sffnvgoe}
\end{figure}
\begin{itemize}
\item  The dip arises from the Fourier transform of the disconnected two-point function. At finite $N$, it gives rise to oscillations due to a finite sum over complex exponentials with integer-separated eigenvalues (on the unfolded scale). More explicitly, the dip as well as the oscillations at finite $N$ are approximated by 
\begin{equation}
Z^2_{\text{dip}}(t)=\frac{\left|\sum _{n=1}^N \ee^{-(\beta+2 \pi \I t) n}\right|^2}{\sum _{n=1}^N \ee^{ -2\beta n}}=
\frac{  \sinh( \beta ) \text{csch} (\beta  N) \bigl(\cos  (2 \pi  N t)-\cosh  (\beta  N)\bigr)}{ \cos(2 \pi  t) -  \cosh(\beta) }~.
\end{equation}
Note that the oscillations are suppressed with increasing $\beta$. Furthermore, the expression above results in recurrences at integer values of $t$ arising from exact in-phase oscillations. However, small deviations away from exact integer eigenvalues results in large cancellations at non-zero $t$ which remove this effect in a finite sample size.
\item The ramp is characterized by a period of nearly linear growth in the spectral form factor.  This linear behavior is a hallmark of large-$N$ random matrix statistics.  In particular, using the exact large-$N$ correlator \eqref{twopt}, one can evaluate the resulting averaged partition functions.  For instance, at $\beta=0$ one has
\begin{equation}\label{eq:Z_infinite_N}
	\left|Z(t)\right|^2=Z^2_{\text{dip}}(t)+\begin{cases}
		2 | t| -| t|  \log (2 | t| +1)\quad&|t|\leq 1~,\\
		2-2 | t|  \coth ^{-1}(2 | t| )&|t|> 1~.
	\end{cases}      
\end{equation}
The expression at finite $\beta$ is easily obtained  by convolving the expression above with $\frac{2 \beta }{\beta ^2+4 \pi ^2 t^2}$ to apply the exponential damping.
A comparison of this result to a finite-$N$ GOE is illustrated in Figure \ref{fig:sffnvgoe}. More generally, one can see that the power-law behavior of the connected correlator $G_{2}(r)-1\sim r^{-2}$ is responsible for the ramp region of the SFF.

\item The plateau marks the transition to the averaged asymptotics of \eqref{lims}. In our conventions, this occurs at time $t=1$.  As explained above, this behavior is characteristic of any theory with a discrete chaotic spectrum.  In the context of RMT, this is reproduced by the fact that the connected two-point function vanishes at small separation, so the large-$t$ behavior is dominated by the disconnected contribution to the correlation function. Note that for the GOE, one expects a smooth transition between the ramp and plateau, as observed in Figure \ref{fig:sffnvgoe}, while for GUE and GSE one instead expects a sharp transition and a peak respectively.
\end{itemize}

\section{Sigma Models: CFTs and Quantum Mechanics}
\label{sigmasec}

As mentioned in the introduction, we will be investigating the spectrum of the Laplacian on certain K\"ahler manifolds with Einstein metrics, which are thus Calabi--Yau in the Ricci-flat case. As we outline in this section, this data has a nice physical interpretation in terms of the spectrum of certain field theories. The theories we consider are sigma models with a non-linear target~\cite{Friedan:1980jf}.  We mostly choose the target to be Calabi--Yau so that the two-dimensional sigma model can be extended to a $(2,2)$ superconformal field theory (SCFT). In practice, however, we will be exclusively focused on the scalar sector in the point-particle limit where the theory simplifies and any Riemannian target is acceptable.

A 2d $(2,2)$ theory can be defined for any compact K\"ahler manifold $M$~\cite{AlvarezGaume:1981hm}.  The defining action for this theory may be written in terms of scalar fields $X^{a}$, viewed as local coordinates on the manifold $M$, as well as fermion fields $\psi^{a}$ and $\lambda^{b}$ which behave as vector fields on the target space.  Schematically, one has
\begin{equation}\label{2dsig}
S=\int_\Sigma \dd^{2}\sigma\left[g_{ab}\partial X^{a}\bar{\partial}X^{b}+g_{ab}(\psi^{a}\bar{D}\psi^{b}+\lambda^{a}D\lambda^{b})+R_{abcd}\psi^{a}\psi^{b}\lambda^{c}\lambda^{d}\right]~,
\end{equation}
where $\sigma^{\mu}=(\tau,\sigma)$ are coordinates on the two-dimensional spacetime $\Sigma$, $g_{ab}(X)$ is a K\"{a}hler metric on the target $M$, and $R_{abcd}(X)$ is the Riemann tensor associated to $g_{ab}$.

Semiclassically, the fields $X^{a}$ are dimensionless and the sigma-model action \eqref{2dsig} defines a conformal field theory for any K\"{a}hler target.  Quantum mechanically, it is well known that there are corrections to this statement and generically conformal invariance is lost~\cite{AlvarezGaume:1980dk,AlvarezGaume:1981hn,AlvarezGaume:1985ww,Gross:1986iv}.  For instance, as an expansion around the large-volume limit of $M$, the first non-trivial contribution to the beta function for the metric $g_{ab}$ is~\cite{Friedan:1980jf}
\begin{equation}\label{betaf}
\beta^g_{ab}= \alpha' R_{ab}+\ldots
\end{equation}
and so conformal invariance is retained at one-loop order only for Ricci-flat targets~\cite{Hull:1985at,AlvarezGaume:1985xfa,Nemeschansky:1986yx}.  In the case of $(2,2)$ supersymmetry where the target is K\"ahler, the vanishing of the one-loop beta function is necessary and sufficient for exact superconformal invariance, since the target is then Calabi--Yau.\footnote{With $(2,2)$ supersymmetry, a Calabi--Yau metric defines an exact SCFT, however the physical metric receives corrections and is generically only K\"ahler to higher-loop order. With $(4,4)$ supersymmetry, as for a K3 target, the physical metric receives no corrections so that the Calabi--Yau metric is the exact target space metric to all loop orders.} 

The $(2,2)$ SCFTs thus obtained are in general strongly interacting and not rational or exactly solvable. They have central charge $c_{L}=c_{R}=\frac{3}{2}\text{dim}(M)$.  Of particular importance to our analysis is that they are known to admit two families of exactly marginal deformations:
\begin{itemize}
\item K\"{a}hler moduli -- these are the size moduli of the Calabi--Yau.  There are $h^{1,1}(M)$ such deformations.  In particular, one of these corresponds to the overall volume, $\text{Vol}(M)$, which scales the overall size of the manifold $M$.
\item Complex structure moduli -- these are the shape moduli of the Calabi--Yau.  There are $h^{2,1}(M)$ such deformations.  For the Calabi--Yau manifolds we encounter below, defined by complex equations in projective space, the complex structure moduli may be viewed as a choice of coefficients in the defining equations modulo redefinition of the coordinates.
\end{itemize}
We will investigate the spectrum of the CFT in the limit of large $\text{Vol}(M)$ when averaged over regions in the complex structure moduli space.  Our interest is in the space of states quantized on a circle, or equivalently the space of local operators. For large volume, the CFT spectrum decomposes into states whose energy remains finite as $\text{Vol}(M)\rightarrow \infty$, known as momentum modes, and solitonic states whose energy diverges in this limit.  The latter are known as winding modes since they arise from geodesics in the target space.\footnote{Modular invariance relates these two classes of states and hence will be lost when we focus on the momentum modes. More discussion of these states in the large-volume limit can be found in \cite{Gao:2013mn}.}  Taking the limit of large volume thus means that we will confine our attention to the momentum modes.  Viewing the two-dimensional spacetime as a string worldsheet, this corresponds to the point-particle limit where the string length is small. In this point-particle limit, the spectrum can be understood in a simple fashion as a supersymmetric quantum mechanics~\cite{Witten:1982df}.\footnote{See also \cite{Figueroa-OFarrill:1997djj,Singleton:2016hky} for further discussion of the map between supersymmetric quantum mechanics and differential forms on the target space.} The Virasoro primaries are labeled by differential forms which are eigenstates of the target-space Laplace operator defined by the metric $g$. Using the K\"{a}hler geometry of the target $M$ we can further decompose these forms by their Hodge type $(p,q)$ with $0\leq p,q\leq \frac{1}{2}\text{dim}(M)$.\footnote{The above describes operators in the NS-NS sector of the $(2,2)$ sigma model.  There are also operators in the Ramond sectors.  These operators begin at a finite gap above the ground state (identity operator) and hence we neglect them in the parametrically large-volume limit where we focus on operators close to $\Delta=0$.}

In terms of the fields appearing in the sigma-model action \eqref{2dsig}, we can understand these operators as follows.  As with differential forms, we can split the fermion fields into those with holomorphic and those with antiholomorphic indices.  Then, given any $(p,q)$-form defined on the target space we can build a operator:
\begin{equation}
\mathcal{O}=\phi_{i_{1}\dots i_{p}\bar{j}_{1}\dots \bar{j}_{q} }(X)\lambda^{i_{1}}\dots \lambda^{i_{p}}\psi^{\bar{j}_{1}}\dots \psi^{\bar{j}_{1}}~.
\end{equation}
This operator is a Virasoro primary if $\phi$ is an eigenform of the Laplace operator, $\Delta \equiv\{\dd,\dd^{\dagger}\}$, where $\dd$ is the exterior derivative operator and $\dd^\dagger$ is its adjoint with respect to $g$.  The scaling dimension $D$ and spin $J$ of $\mathcal{O}$ are then given by
\begin{equation}\label{eq:scaling}
D=\Delta+\frac{p+q}{2}~, \hspace{.5in}J=\frac{p-q}{2}~.  
\end{equation}
Thus the problem of calculating the spectrum of primaries in the large-volume limit is reduced to studying the Laplace operator on the sigma-model target space.

In what follows we will make a further simplification by truncating to those operators constructed only out of the scalar operators $X^a$.  This means that we are studying the Laplace operator acting on functions (as opposed to forms).  We can deduce when this approximation is self consistent by applying Weyl's law for the asymptotics of eigenvalues~\cite{Weyl11}.  Specifically this states that for large $n$, the $n$-th eigenvalue of the Laplacian, $\lambda_{n}$, is approximately
\begin{equation}\label{weyl}
\lambda_{n}\xrightarrow{n\rightarrow \infty}4\pi \left(\frac{\Gamma(1+d/2)n}{\Vol(M)}\right)^{2/d}~,
\end{equation}
where $d$ is the dimension of the target space $M$.  We want the primary operator associated to the $n$-th eigenvalue to have scaling dimension much smaller than those primaries that come from the forms we neglect.  In particular, this means that all the operators we consider are scalars with scaling dimensions much less than one.  We can still take $n$, the number of operators, large provided that we simultaneously take the volume of $M$ large:
\begin{equation}\label{volnf}
n \ll \frac{\Vol(M)}{(4\pi)^{d/2}\Gamma(1+d/2)}~.
\end{equation}
In the following we always assume that the volume is chosen to satisfy this constraint.  Since we mostly examine the statistics of the unfolded spectrum, the numerical value of the scaling dimension is not relevant, and we often set it to one for convenience.  In practice, one can always restore the absolute dimensions by scaling the eigenvalues appropriately.

In summary, in the large-volume limit we can consistently truncate to studying the primary operators of the bosonic sigma model of a scalar field $X$ valued in the target $M$.  The spectrum of Virasoro primaries is then given by the eigenfunctions of the Laplace operator, and the CFT partition function is
\begin{equation}
Z(\tau,\bar{\tau})=\frac{1}{|\eta(\tau)|^{2}}\sum_{n}\exp(-\beta \lambda_{n})~.
\end{equation}
Up to the factors of $\eta(\tau)$ capturing the conformal descendants, this is the same as the quantum mechanical problem of studying a (bosonic) point particle moving on $M$. Our goal is now to compute this spectrum and to compare its properties when averaged over complex structure moduli to that of RMT.

\section{The Laplacian}\label{laplacian}

As we have outlined, we are interested in computing the spectrum of the scalar Laplacian on compact K\"ahler manifolds with Ricci-flat metrics. In this section, we give a general overview of the Laplacian and a discussion of our numerical methods for finding both the metric and the spectrum.

\subsection{The Laplacian}

Given a metric $g$ on a closed $d$-dimensional manifold $M$, the scalar Laplacian is a real differential operator given by
\begin{equation}
\Delta=\dd^{\dagger}\dd~,
\end{equation}
where $\dd^{\dagger}=-\star\op d\star$ is the adjoint of $\dd$ with respect to the inner product $\langle\,,\,\rangle$ defined by the metric $g$, and $\star$ is the Hodge star operator. Since $\star$ depends on the metric, the spectrum of $\Delta$ depends on the choice of metric. The spectrum of the Laplacian is then the set of eigenvalues of $\Delta$ acting on the space of functions, where an eigenfunction $\phi$ has eigenvalue $\lambda$ defined by
\begin{equation}
\Delta\phi=\lambda\phi~.\label{eq:eigenvalue_equation}
\end{equation}
The Laplacian is a self-adjoint operator with respect to the inner product and so its eigenvalues are real. The eigenvalues are also non-negative as can be seen from
\begin{equation}
\lambda=\frac{\langle\phi,\Delta\phi\rangle}{\langle\phi,\phi\rangle}=\frac{\langle\dd\phi,\dd\phi\rangle}{\langle\phi,\phi\rangle}=\frac{\Vert\dd\phi\Vert^{2}}{\Vert\phi\Vert^2}\geq 0~.
\end{equation}
Eigenfunctions with different eigenvalues are orthogonal with respect to the inner product, so that they can be normalized to $\langle\phi_{m},\phi_{n}\rangle=\delta_{mn}$. Moreover, the eigenfunctions form a complete basis for the space of square-integrable functions, so that the product of any two eigenfunctions can be written as a sum over the basis (reminiscent of the operator product expansion for chiral fields).

Eigenfunctions with eigenvalue zero are known as zero-modes or harmonic functions. On a compact connected manifold there is only one harmonic function (up to an overall scale) given by the constant function, and so there will be a single zero-mode. Furthermore, compactness also implies that the spectrum is discrete and that the eigenspaces -- spanned by eigenfunctions with the same eigenvalue -- are finite dimensional. If the metric admits a non-abelian isometry group (continuous or discrete), the eigenspaces can have dimensions greater than one and the corresponding eigenvalues appear with some multiplicity.  In the examples we will consider, the metrics we will be somewhat generic and so their isometry groups will be trivial. Thanks to this, the eigenvalues will all be distinct with multiplicity equal to one. Due to the metric dependence of the Laplacian, the eigenvalues of $\Delta$ scale as $\lambda\sim g^{ab}$. In terms of the volume $\op{Vol}(M)\equiv\int_{M}\vol=\int_{M}\star1$, the eigenvalues scale as
\begin{equation}
\lambda\sim\op{Vol}(M)^{-2/d}~.
\end{equation}
In all examples, we scale the metric so that $\op{Vol}(M)=1$.

Given a choice of basis $\{\alpha_{A}\}_{\infty}$ for the (infinite-dimensional) space of complex functions on $M$, the eigenvalues and eigenfunctions of the Laplacian can be computed from the matrix elements of $\Delta$. Expanding a given eigenfunction as $\phi=\phi^{A}\alpha_{A}$, where $\phi^{A}$ is a vector of complex numbers, the eigenvalue equation (\ref{eq:eigenvalue_equation}) is equivalent to
\begin{equation}
\Delta_{AB}\phi^{B}=\lambda O_{AB}\phi^{B}~,\label{eq:gen_eigen}
\end{equation}
where $\Delta_{AB}\equiv\langle\alpha_{A},\Delta\alpha_{B}\rangle$ are the matrix elements of the Laplacian and $O_{AB}\equiv\langle\alpha_{A},\alpha_{B}\rangle$ encodes the non-orthonormality of the chosen basis. The eigenvalues and eigenvectors for this ``generalized eigenvalue problem'' then give the spectrum of the Laplacian and the expansion of the eigenfunctions in the basis $\{\alpha_{A}\}_{\infty}$. 

Though there are many results on the spectrum of Laplace-type operators on Riemannian manifolds~\cite{EigenvaluesinRiemannianGeometry,Canzani13}, there are few sharp results concerning the eigenvalues and eigenfunctions of the Laplacian for an arbitrary metric. Since zero-modes are harmonic, they have a cohomological interpretation and their multiplicities are given by $|\pi_{0}(X)|$, the number of connected components of $M$. The massive-modes -- eigenfunctions with $\lambda>0$ -- have no such interpretation and so are much less well understood (though see \cite{1910.04767,2007.10337,Bonifacio:2021msa} for recent results on using consistency conditions to prove non-trivial ``bootstrap'' bounds on both eigenvalues and triple overlap integrals). Even if the metric on $M$ is known analytically, it is usually not possible to determine either the eigenvalues or eigenfunctions explicitly, forcing one to resort to numerical methods. Furthermore, there are cases where the metric itself is not known explicitly and so must also be determined numerically -- this is the case for the metrics that we are interested in.

As we mentioned around \eqref{weyl}, one of the few results for the spectrum of the Laplacian on a generic Riemannian manifold is Weyl's law, which gives an asymptotic expression for the number of eigenvalues $N(\lambda)$ less than $\lambda$:
\begin{equation}\label{weyl_N}
	N(\lambda) = \frac{\Vol(M)}{(4\pi)^{d/2}\Gamma(1+d/2)}\lambda^{d/2}~.
\end{equation}
Since $N(\lambda)=\int\dd\lambda\,\rho(\lambda)$, the asymptotic eigenvalue density $\rho(\lambda)$ is thus
\begin{equation}\label{weyl_density}
	\rho(\lambda) =  \frac{\Vol(M)}{(4\pi)^{d/2}\Gamma(d/2)}\lambda^{d/2-1}~.
\end{equation}

At this point we further restrict to manifolds which admit a Kähler structure. This means $M$ is even-dimensional and the metric is hermitian. Furthermore, $M$ admits a complex structure that is compatible with the metric, and the exterior derivative then decomposes in terms of the Dolbeault operators as $\dd=\partial+\bar{\partial}$. Finally, $M$ admits a non-degenerate real two-form $\omega$ known as the Kähler form. Both the Kähler form and the metric are then determined by a choice of locally defined real function $K$ known as a Kähler potential. The examples we will consider will actually be Kähler--Einstein (KE), so that the Ricci tensor is proportional to the metric.\footnote{See \cite{EinsteinManifolds} and references therein for a discussion of Einstein manifolds.} Explicit non-trivial examples of such metrics are few and far between, especially for Ricci-flat metrics. Instead, we will use a numerical method to calculate an approximate metric and then compute the matrix elements of the Laplacian to determine the spectrum.

\subsection{Numerical Metrics and the Spectrum of \texorpdfstring{$\Delta$}{Delta}}

We now review our numerical approach for computing both metrics and the spectrum of the Laplacian. Many detailed discussions of both the underlying mathematics and its practical implementation have appeared in the literature~\cite{math/0512625,hep-th/0506129,hep-th/0612075,0712.3563,0908.2635,1910.08605,Cui:2019uhy,Ashmore:2020ujw,2012.04656,2012.04797,2012.15821}, so we shall give only a brief outline. 

Beginning with the metric, the rough idea is to approximate the Kähler potential of the metric using a finite-dimensional basis of degree-$k_g$ polynomials on $M$, where the precise combination of polynomials is parametrized by a hermitian matrix $h^{\alpha\bar{\beta}}$ of constants. As $h^{\alpha\bar{\beta}}$ is varied, the Kähler potential describes different metrics within the same Kähler class. The goal is then to pick the parameters to find a good approximation to the desired metric, where goodness is measured by how close to Ricci-flat the resulting metric is.\footnote{Though our discussion focuses on Ricci-flat metrics, the same applies to Einstein metrics.} One way to fix these parameters is via an iterative procedure proposed by Donaldson whose fixed point is a ``balanced'' metric~\cite{math/0512625}. As $k_g$ is increased, the size of the basis of polynomials is increased and the balanced metric is a better approximation. In the limit $k_g\to\infty$, the balanced metric is unique and converges to the exact K\"ahler--Einstein or Ricci-flat metric.

All of the examples we consider are hypersurfaces defined by the vanishing locus of some holomorphic function in an ambient projective space. To be concrete, let us focus on the example where $M$ is a quintic threefold. The defining equation is a quintic equation in $\bP^{4}$
\begin{equation}\label{eq:def_eq}
f(z)=\sum_{m,n,p,q,r}c_{mnpqr}z_{m}z_{n}z_{p}z_{q}z_{r}~,
\end{equation}
where $[z_{0}:z_{1}:z_{2}:z_{3}:z_{4}]$ are homogeneous coordinates on the projective space and $c_{mnpqr}$ are complex constants that parametrize the complex structure moduli of the quintic. Taking $k_g$ to be a positive integer, we let $\{s_{\alpha}\}$ be a basis for $H^{0}(M,\mathcal{O}_{M}(k))$, where $\alpha=1,\ldots,N_{k}$ and $N_{k}=\dim H^{0}(M,\mathcal{O}_{M}(k))$. This basis can be chosen to be the degree-$k_g$ holomorphic monomials on $\bP^{4}$ modulo $f=0$. We then make an ansatz for the Kähler potential as
\begin{equation}
K=\frac{1}{\pi k_g}\ln\bigl(s_{\alpha}h^{\alpha\bar{\beta}}\bar{s}_{\bar{\beta}}\bigr)~,\label{eq:kahler_potential}
\end{equation}
where $h^{\alpha\bar{\beta}}$ is a constant hermitian matrix. This gives an ``algebraic metric'', a simple generalization of the usual Fubini--Study Kähler potential. Note that linear combinations of the functions $s_{\alpha}\bar{s}_{\bar{\beta}}$ at degree $k_g$ give the eigenfunctions for the first $k_g+1$ eigenspaces of the Laplacian on $\bP^{4}$, so that (\ref{eq:kahler_potential}) can be interpreted as a spectral expansion with coefficients $h^{\alpha\bar{\beta}}$~\cite{math/0512625,0908.2635}. 

The iterative procedure which fixes $h^{\alpha\bar{\beta}}$ starts with Donaldson's ``$T$-operator'':\footnote{Here $\vol$ (which defines $\op{Vol}(X)$) is some volume form on $X$. In the Calabi--Yau case, since the holomorphic $d/2$-form $\Omega$ can be determined exactly using a residue theorem, the volume form can be chosen to be the exact Calabi--Yau measure defined by $\Omega\wedge\bar{\Omega}$. Otherwise, one can choose the volume to be that defined by the numerical metric computed from $h^{\alpha\bar\beta}$ at the previous iteration.}
\begin{equation}\label{eq:T_op}
T\colon h^{\alpha\bar{\beta}}\mapsto T(h)_{\alpha\bar{\beta}}=\frac{N_{k}}{\op{Vol}(M)}\int_{M}\vol\frac{s_{\alpha}\bar{s}_{\bar{\beta}}}{s_{\gamma}h^{\gamma\bar{\delta}}\bar{s}_{\bar{\delta}}}~.
\end{equation}
As $n\to\infty$, the sequence $h_{(n+1)}=T(h_{(n)})^{-1}$ converges to give the balanced metric $g_{i\bar{j}}=\partial_{i}\partial_{\bar{j}}K$ at degree $k_g$.\footnote{We are using $i,j$ to label complex coordinates on $M$ while $a,b$ are real coordinates. Homogeneous coordinates on projective space are labeled with $m,n,\dots$} The resulting metric is automatically Kähler (since it comes from a Kähler potential) and gives an approximation to the Kähler--Einstein metric on $M$, which in this example is Ricci-flat and so Calabi--Yau. Note that the K\"ahler class of the resulting metric is given by $c_1(\mathcal{L})$, where $\mathcal{L}=\mathcal{O}_M(1)$ in our example. If $h^{1,1}(M)>1$, then one could choose a different K\"ahler class (or more precisely a different ray in the K\"ahler cone) by choosing $\mathcal{L}$ appropriately.

The integrals over $M$ have to be evaluated using a Monte Carlo method by sampling points on $M$ and then summing over them. In the examples that follow, the points were generated using the intersecting lines method laid out in~\cite{hep-th/0612075,0712.3563}. The points were then resampled as in ``sequential Monte Carlo'' to ensure that regions of $M$ are not undersampled. We will denote the number of points used for these numerical integrals by $N_{p}$ in what follows. Furthermore, since \eqref{eq:T_op} converges rather quickly in practice, we use 20 iterations to get sufficiently close to the balanced metric.

Given an approximate metric $g_{i\bar{j}}$ computed as above, one can calculate the spectrum of the Laplacian by evaluating the matrix elements of $\Delta$ in some convenient basis. Again, we keep our discussion brief -- more details can be found in~\cite{0804.4555,0805.3689,Ashmore:2020ujw,2103.07472}. As mentioned earlier, in practice we have to restrict to some finite-dimensional basis which approximates the space of functions in a controlled way. We denote this basis by $\{\alpha_{A}\}$. As with the Kähler potential, it is natural to use the eigenfunctions of the Laplacian on the ambient space:
\begin{equation}
\{\alpha_{A}\}=\frac{\{s_{\alpha}^{(k_{\phi})}\bar{s}_{\bar{\beta}}^{(k_{\phi})}\}}{\bigl(|z_{0}|^{2}+\ldots+|z_{4}|^{2}\bigr)^{k_{\phi}}}~,
\end{equation}
where $k_{\phi}$ is a non-negative integer and $\{s_{\alpha}^{(k_{\phi})}\}$ is a basis for $H^{0}(M,\mathcal{O}_{M}(k_{\phi}))$. Here the denominator ensures that the functions do not transform under rescaling of the homogeneous coordinates and so are well defined on $\bP^{4}$. Increasing $k_{\phi}$ increases the size of the approximate basis so that the resulting matrix elements $\Delta_{AB}$ of the Laplacian define a larger matrix. The effect of this is two-fold: one can better approximate the exact eigenfunctions and eigenvalues of $\Delta$, and one can also compute more of them (since a larger matrix has more eigenvalues). In practice, one wants to compute the spectrum with as large a basis as is computationally feasible and then drop the upper end of the spectrum, since it is those eigenvalues that will be most affected by the finite size of the approximate basis.

Given a numerical metric and a choice of a basis $\{\alpha_{A}\}$, one can then numerically evaluate the matrices that appear in (\ref{eq:gen_eigen}). With these in hand, it is simple to solve for the eigenvalues and eigenvectors, which determine the spectrum of $\Delta$ and its eigenfunctions respectively.\footnote{The code for the both the metric and the Laplacian was written in \emph{Mathematica}~\cite{Mathematica}. More details of the implementation can be found in \cite{Ashmore:2020ujw}.}

\section{Results for Spectral Statistics}\label{eq:results}

In this section, we summarize our results for the average statistical properties of an ensemble of sigma model CFTs.  We carry out this analysis for a genus-three curve, a K3 complex surface, and a quintic Calabi--Yau threefold.  (Although the genus-3 curve does not yield an exact (2,2) CFT due to its curvature, it is a useful warm-up to our general analysis and also exhibits RMT statistics in its spectrum.)

In each case we work in the region of large volume as detailed in Section \ref{sigmasec}.  We compute the spectrum of the Laplacian for 1,000 random choices of complex structure, corresponding to marginal couplings in the CFT, and then average over the resulting ensemble.  We give the eigenvalue density in each case (without unfolding), which via \eqref{eq:scaling} gives the spectrum of scaling dimensions of scalar operators in the corresponding sigma model. For comparison, we also include the asymptotic cumulative density predicted by Weyl's law from \eqref{weyl_N}. Finally, we unfold the spectrum as described in Section \ref{rmtsec} and compare our results to the expected spectral statistics of random matrix theory, finding good agreement with GOE in all cases.

Before we continue, we should make a comment on how the complex structures are chosen. In all cases, the examples we consider are defined by the zero locus of a holomorphic equation in some complex projective space. The choice of complex structure corresponds to the choice of coefficients in these equations, for example the $c_{mnpqr}$ in \eqref{eq:def_eq}. In this paper, we always sample these coefficients from the unit disk in the complex plane randomly with respect to a flat measure. Obviously, there are other measures that one could choose. For example, when averaging over the moduli space of 2d CFTs with a toroidal target space, such as in \cite{Afkhami-Jeddi:2020ezh,Maloney:2020nni,Benjamin:2021wzr}, it is natural to take the measure to be the one induced by the metric on the moduli space of the torus. This agrees with both the Zamolodchikov metric in the field theory and the Weil--Petersson metric in the geometry. Extending these ideas to higher-dimensional Calabi--Yau targets, it would seem natural to use again the Weil--Petersson metric on the moduli space of complex structures to define the measure. In principle, these metrics can be obtained via a combination of mirror symmetry and localization~\cite{Jockers:2012dk,1206.2606,1210.6022}, by an analysis of their special geometry~\cite{Aleshkin:2017oak,Aleshkin:2017fuz,Aleshkin:2018jql}, or numerically~\cite{Keller:2009vj}. Unfortunately, for high-dimensional moduli spaces, such as that of the quintic, these calculations quickly become unfeasible, and so we will simply work with a flat measure. From previous experience with these metrics~\cite{Candelas:1990rm,1803.04989,1905.05225,2103.07472}, they are usually rather flat away from singular points in moduli space. Since such points are a measure zero set, random sampling is likely to avoid them and instead simply sample the relatively flat regions. Due to this, we do not believe including the moduli space metric will qualitatively change our results.\footnote{However, the metric would be essential if one wanted to average over the \emph{entire} moduli space.}

\subsection{Genus-3 Curves}

We start by considering a simple example which is not Calabi--Yau, namely a genus-3 Riemann surface or complex curve. These surfaces admit K\"ahler metrics of constant negative curvature, and so they are K\"ahler--Einstein. Looking back at the  beta function in \eqref{betaf}, a sigma model with a genus-3 target equipped with such a KE metric will not be asymptotically free in the UV, but can be made weakly coupled in the IR.\footnote{Of course, like all 2d sigma models, this theory is conformal at tree level and the violation of scale invariance occurs only through quantum corrections.}

A genus-3 curve can be defined as the vanishing locus of a quartic equation in $\bP^{2}$
\begin{equation}
f=\sum_{m,n,p,q}c_{mnpq}z_{m}z_{n}z_{p}z_{q}\ ,\label{eq:g3_curve}
\end{equation}
where the $c_{mnpq}$ are chosen randomly from the unit disk in the complex plane with a flat measure. There are $\binom{3+4-1}{4}=15$ independent components in the $c_{mnpq}$ and 9 of these can be absorbed using $\GL{3,\bC}$ transformations of the homogeneous coordinates. This leaves 6 degrees of freedom which are simply the complex structure moduli that parametrize a six-dimensional family of curves.\footnote{Constant curvature metrics on a genus-$\gamma$ curve are specified by $6\gamma-6$ parameters, so the genus-3 curves described by (\ref{eq:g3_curve}) give a six-dimensional subspace of the full 12-dimensional moduli space.} The spectrum of the Laplacian on such hyperbolic surfaces cannot be computed explicitly, and so one must use numerics. We use Donaldson's algorithm to find an approximate KE metric on the curve and then compute the spectrum numerically.\footnote{Alternative approaches for computing the spectrum were proposed in \cite{1110.2150,Cook18}. In particular, our numerical method reproduces the spectra of the Klein and Fermat quartics in $\mathbb{P}^2$ given in \cite[Appendix D]{Cook18}.}

We computed the spectrum for 1,000 samples, giving us an ensemble of eigenvalues whose statistics we can examine. For each sample, the approximate K\"{a}hler--Einstein metric was computed at $k_g=6$ with 20 iterations of the $T$-operator.  In each case, the spectrum of the Laplacian was then computed at $k_{\phi}=3$, allowing calculation of the first 324 eigenvalues -- of these, we keep the lowest-lying 160 to avoid numerical errors at the edge of the computed spectrum. For both calculations $N_{p}=10^{6}$ points were used for numerical integrations.

\begin{figure}
	\includegraphics[width=2.49in]{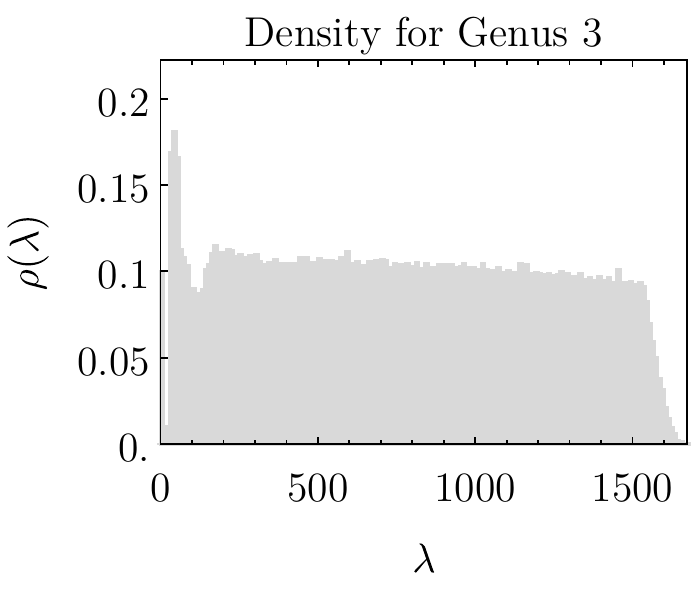}\qquad\includegraphics[width=2.49in]{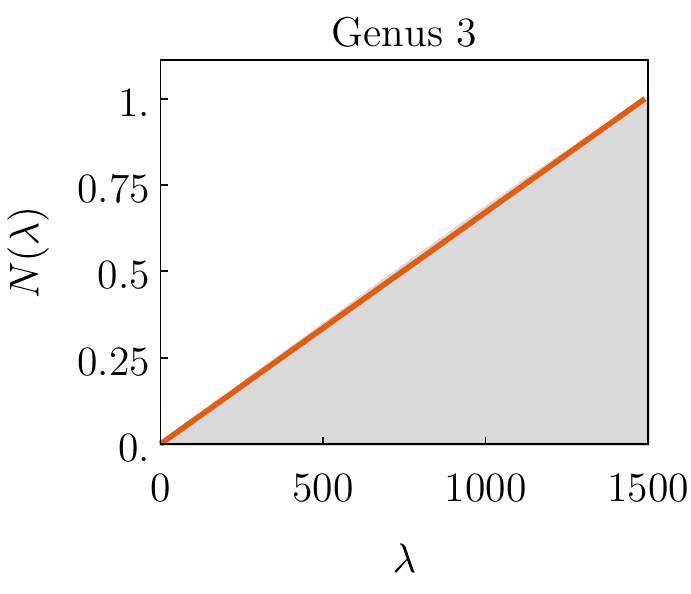}\\
	\caption{On the left, we show the eigenvalue density for the Laplacian on genus-3 curves drawn from an ensemble of 1,000 samples in complex structure moduli space with each sample containing the lowest-lying 160 eigenvalues. The peak at $\lambda$ is due to the zero-mode which is present in every instance of the ensemble. On the right, we show the integrated eigenvalue density up to $\lambda=1500$ together with the asymptotic number of eigenvalues predicted by Weyl's law \eqref{weyl_N} in orange.}
	\label{fig:rho_g3}
\end{figure}

The resulting eigenvalue density for the ensemble is shown in Figure \ref{fig:rho_g3}. The density is relatively flat, which is somewhat expected since Weyl's law in two dimensions predicts that the asymptotic density is constant. Note that the sudden decrease in the density at around $\lambda=1600$ is an artifact of keeping only the first 160 eigenvalues for each sample. 

\begin{figure}
	\includegraphics{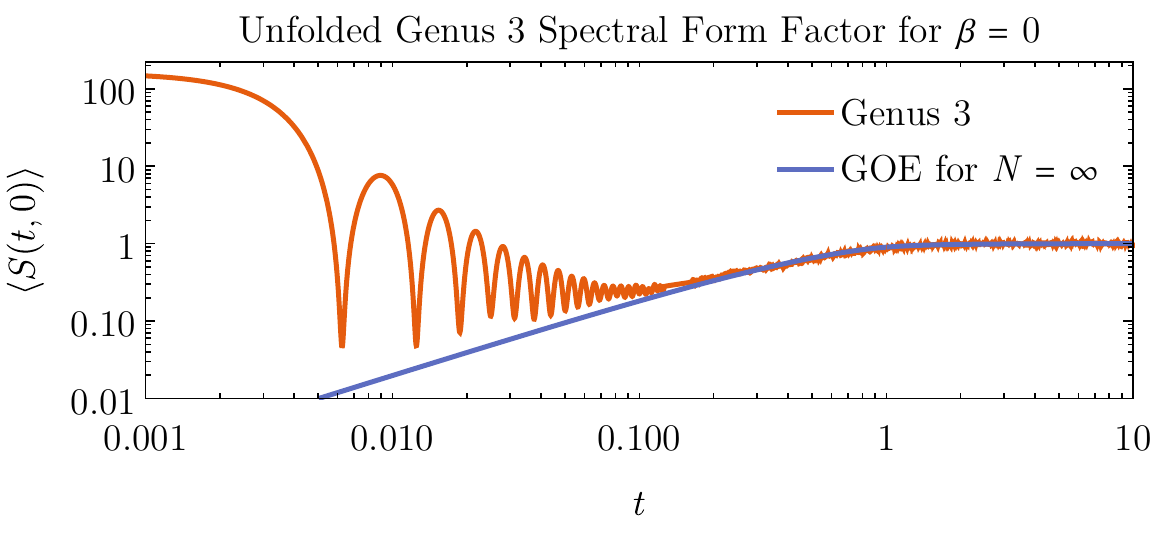}\\
	\includegraphics{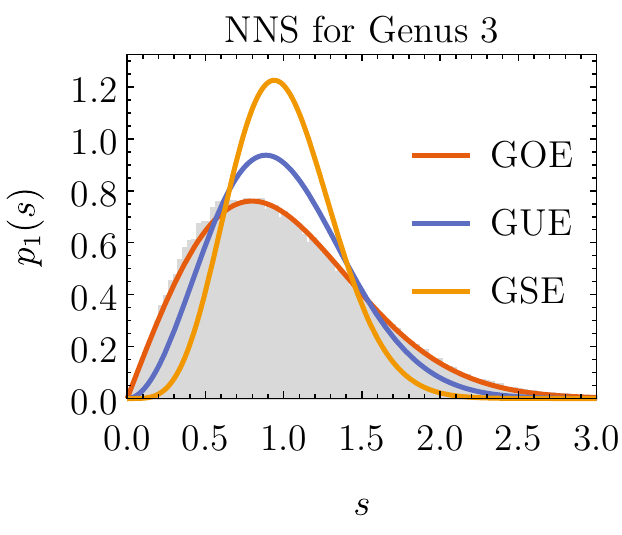}\qquad\includegraphics{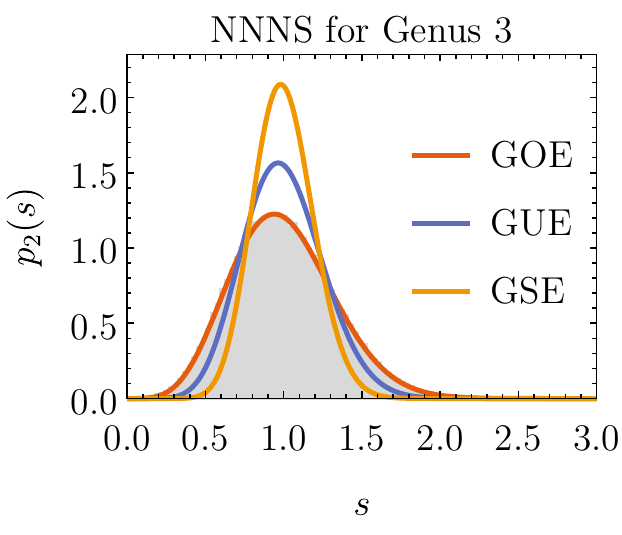}\\
	\includegraphics{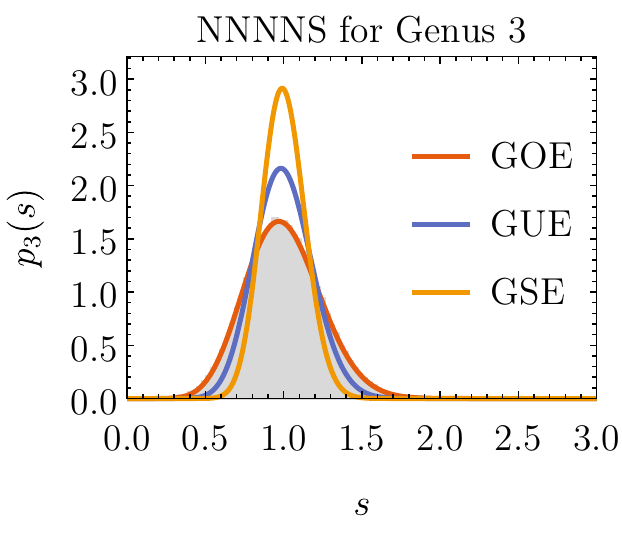}\qquad\includegraphics{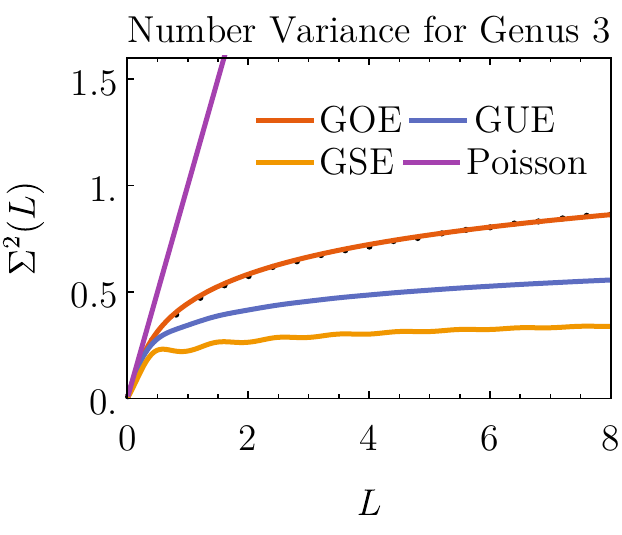}
	\caption{The top row shows the unfolded spectral form factor for $\beta=0$ for genus-3 curves. The orange curve shows an ensemble average consisting of 1,000 samples in complex structure moduli space with each sample containing the lowest-lying 160 eigenvalues. The dip, ramp and plateau are all present. The second and third rows display certain spectral statistics together with fits to GOE.  (Other matrix ensembles are also shown for comparison.) The second row shows the nearest-neighbor level spacings (NNS) and the next-to-nearest (NNNS). The third row shows the next-to-next-to-nearest (NNNNS) level spacings and the number variance.}
	\label{fig:g3}
\end{figure}

After unfolding the spectrum as described around \eqref{eq:unfold}, the spectral statistics can be compared to a random matrix ensemble. This is illustrated in Figure \ref{fig:g3}, where we find a good fit to the GOE. In more detail: the unfolded SFF displays the dip, ramp and plateau we expect from GOE, and the next$^{k-1}$ nearest-neighbor spacings for $k=1,2,3$ and the number variance match that of a GOE. Note that it is known that Riemann surfaces with constant negative curvature metrics admit ergodic geodesic flows~\cite{Hedlund,Hopf2,Hopf1}, and so GOE statistics in the spectrum is expected. However, to the best of our knowledge, there is still no proof of this (see, for example, \cite[\S9.13.2]{Berger}). Our results support this conjecture.

\subsection{K3 Complex Surfaces}

A K3 surface can be defined as the vanishing locus of a quartic equation in $\bP^{3}$
\begin{equation}
f=\sum_{m,n,p,q}c_{mnpq}z_{m}z_{n}z_{p}z_{q}\ ,
\end{equation}
where the $c_{mnpq}$ are chosen randomly from the unit disk in the complex plane with a flat measure. There are $\binom{4+4-1}{4}=35$ independent components in the $c_{mnpq}$ and 16 of these can be absorbed using $\GL{4,\bC}$ transformations of the homogeneous coordinates. This leaves 19 degrees of freedom which can be identified with the complex structure moduli that parametrize the family of smooth quartic K3 surfaces.\footnote{The space of complex analytic K3 surfaces is 20-dimensional, with algebraic quartic K3 surfaces filling out a 19-dimensional subspace~\cite{hep-th/9611137}.}

Again we computed 1,000 samples. For each sample, the approximate Ricci-flat metric was computed at $k_g=6$ with 20 iterations of the $T$-operator.  The spectrum of the Laplacian was computed at $k_{\phi}=3$, allowing calculation of the first 400 eigenvalues -- of these, we keep the lowest-lying 200. For both calculations, $N_{p}=10^{6}$ points were used for numerical integrations. 

\begin{figure}
	\includegraphics[width=2.49in]{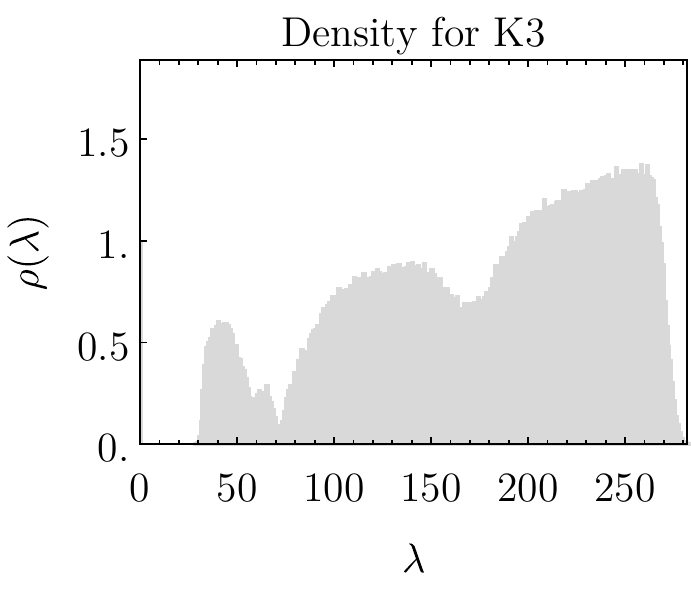}\qquad\includegraphics[width=2.49in]{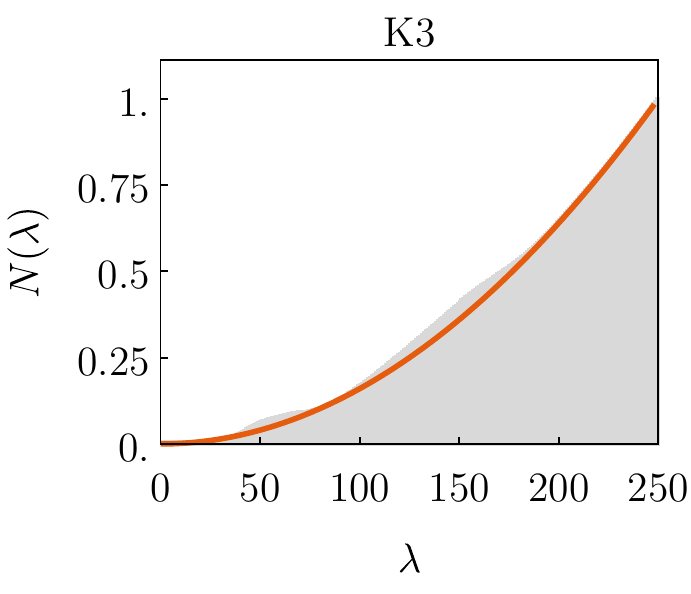}\\
	\caption{On the left, we show the eigenvalue density for the Laplacian on K3 surfaces drawn from an ensemble of 1,000 samples in complex structure moduli space with each sample containing the lowest-lying 200 eigenvalues. On the right, we show the integrated eigenvalue density up to $\lambda=250$ together with the asymptotic number of eigenvalues predicted by Weyl's law \eqref{weyl_N} in orange.}
	\label{fig:rho_K3}
\end{figure}

We show the resulting eigenvalue density for the ensemble in Figure \ref{fig:rho_K3}. The density is much less regular compared to the genus-3 curve. This might be explained by the fact that there are 19 complex structure moduli for the K3, as opposed to 6 for the genus-3 case -- 1,000 points captures much less of the 19-dimensional space than the 6-dimensional one. One might imagine that greatly increasing the size of the ensemble would even out some of the peaks and troughs that are present in the density. Again, we note that the sudden drop in the density at around $\lambda=275$ is an artifact of keeping only the first 200 eigenvalues for each sample.

\begin{figure}
	\includegraphics{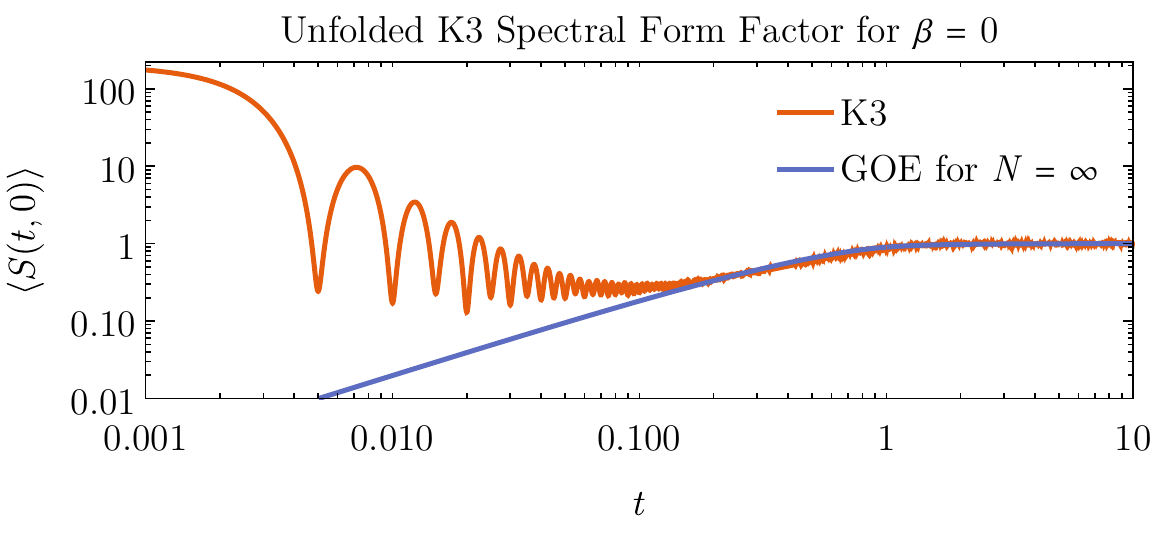}\\
	\includegraphics{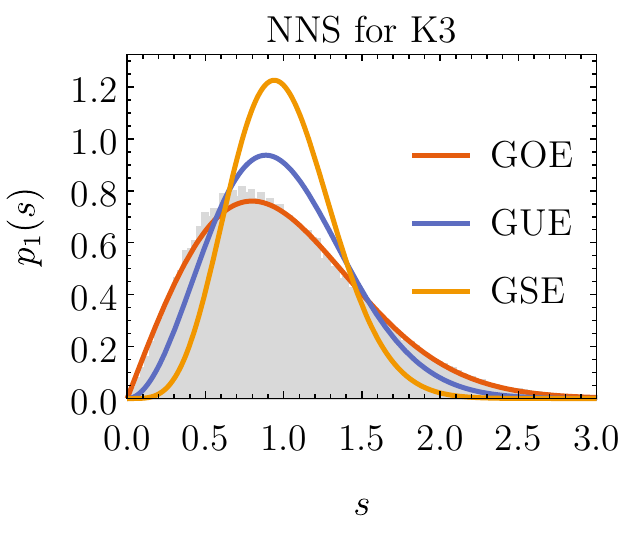}\qquad\includegraphics{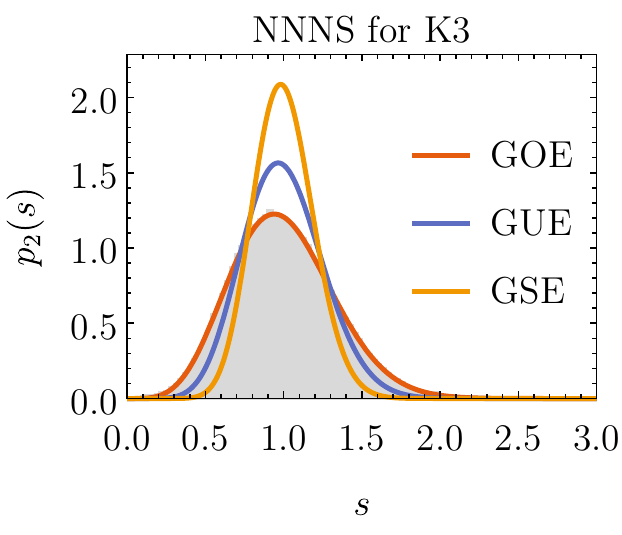}\\
	\includegraphics{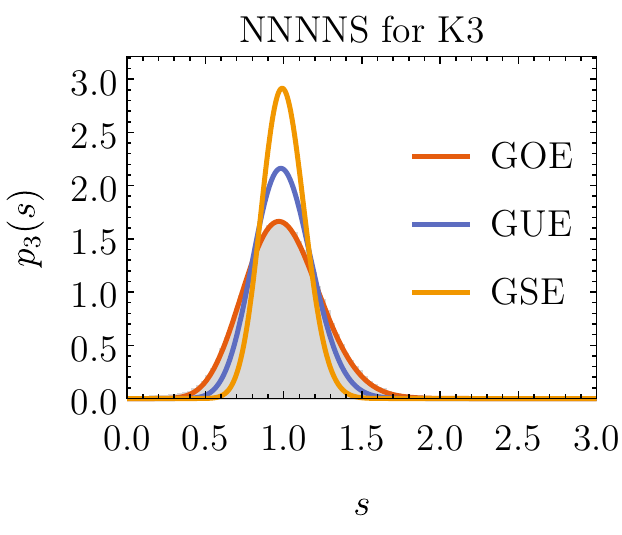}\qquad\includegraphics{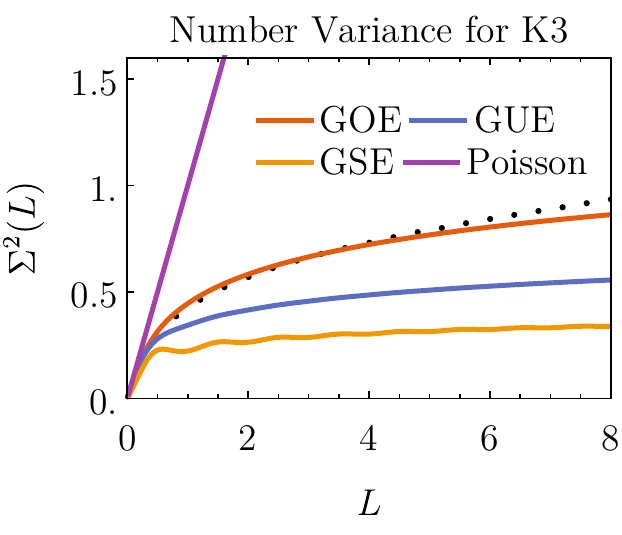}
	\caption{The top row shows the unfolded spectral form factor for $\beta=0$ for K3 surfaces. The orange curve shows an ensemble average consisting of 1,000 samples in complex structure moduli space with each sample containing the lowest-lying 200 eigenvalues. The dip, ramp and plateau are all present. The second and third rows display certain spectral statistics together with fits to GOE.  (Other matrix ensembles are also shown for comparison.) The second row shows the nearest-neighbor level spacings (NNS) and the next-to-nearest (NNNS). The third row shows the next-to-next-to-nearest (NNNNS) level spacings and the number variance.}
	\label{fig:K3}
\end{figure}

Again, we unfold the spectrum, as described around \eqref{eq:unfold}, and compare the spectral statistics to a random matrix ensemble. The plots of the SFF, nearest-neighbor spacings and number variance are shown in Figure \ref{fig:K3}. As with the genus-3 case, we find a good fit to the GOE for both short- and long-range statistics.

\subsection{Quintic Calabi--Yau Threefolds}

A quintic threefold can be defined as the vanishing locus of a quintic equation in $\bP^{4}$
\begin{equation}
f=\sum_{m,n,p,q,r}c_{mnpqr}z_{m}z_{n}z_{p}z_{q}z_{r}\ ,
\end{equation}
where the $c_{mnpqr}$ are chosen randomly from the unit disk in the complex plane with a flat measure. There are $\binom{5-1}{5}=126$ independent components in the $c_{mnpqr}$ and 25 of these can be absorbed using $\GL{5,\bC}$ transformations of the homogeneous coordinates. This leaves 101 degrees of freedom which can be identified with the complex structure moduli that parametrise the family of smooth quintic threefolds.

For 1,000 samples, the approximate Ricci-flat metric was computed at $k_g=6$ with 20 iterations of the $T$-operator.  The spectrum of the Laplacian was then computed at $k_{\phi}=2$, allowing calculation of the first 225 eigenvalues -- of these, we keep the lowest-lying 100. For both calculations, $N_{p}=1.2\times10^{6}$ points were used for numerical integrations.

\begin{figure}
	\includegraphics[width=2.49in]{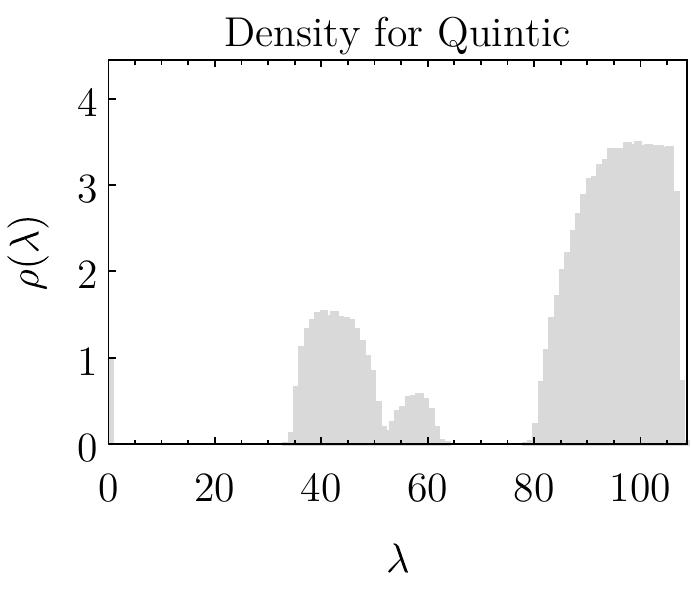}\qquad\includegraphics[width=2.49in]{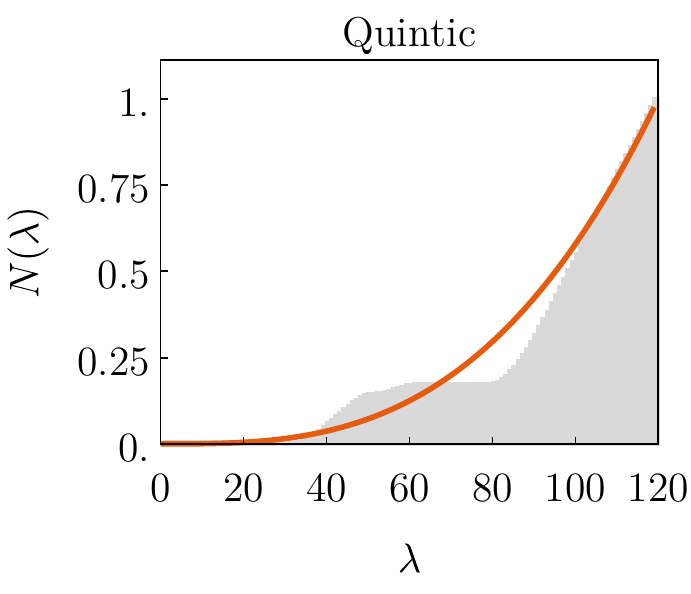}\\
	\caption{On the left, we show the eigenvalue density for the Laplacian on quintic Calabi--Yau threefolds drawn from an ensemble of 1,000 samples in complex structure moduli space with each sample containing the lowest-lying 100 eigenvalues. On the right, we show the integrated eigenvalue density up to $\lambda=120$ together with the asymptotic number of eigenvalues predicted by Weyl's law \eqref{weyl_N} in orange -- note that to more easily see the Weyl's law behavior, we have used the lowest-lying 200 eigenvalues in this plot.}
	\label{fig:rho_QF}
\end{figure}

We show the resulting eigenvalue density for the ensemble in Figure \ref{fig:rho_QF}. The density is much less regular than both the genus-3 and K3 cases, with pronounced gaps in the density where $\rho(\lambda)=0$. Since a quintic threefold has a 101-dimensional space of complex structures, 1,000 points probe very little of the space. One would likely need a very large ensemble to even out some of the peaks and troughs present in the density. Note that the results of \cite{2103.07472} suggest that such a smoothing out of the density could indeed occur: there it was found that taking a single complex structure parameter to be large caused some eigenvalues to decrease in size, while others increased. Again, we note that the sudden drop in the density at around $\lambda=110$ is an artifact of keeping only the first 100 eigenvalues for each sample.

After unfolding the spectrum, we compare the spectral statistics to a random matrix ensemble in Figure \ref{fig:QF}. As with both the genus-3 and K3 examples, we find a good fit to the GOE for both short- and long-range statistics.

\section*{Acknowledgments}

We thank L.~Delacretaz and E.~Mazenc for useful discussions. AA is supported by the European Union's Horizon 2020 research and innovation program under the Marie Sk\l{}odowska-Curie grant agreement No.\,838776. The work of CC and NAJ is supported by the US Department of Energy DE-SC0021432. This work was completed in part with resources provided by the University of Chicago Research Computing Center. 

\bibliographystyle{utphys.bst}
\bibliography{biblio,citations_inspire}

\providecommand{\href}[2]{#2}\begingroup\raggedright\begin{thebibliography}{100}

\bibitem{mehta2004random}
M.~Mehta, \href{http://dx.doi.org/10.1016/C2009-0-22297-5}{{\em Random
  Matrices}}, vol.~142 of {\em Pure and Applied Mathematics}.
\newblock Elsevier/Academic Press, 2004.

\bibitem{Br_zin_1997}
E.~Br\'ezin and S.~Hikami, ``Spectral form factor in a random matrix theory'',
  \href{http://dx.doi.org/10.1103/PhysRevE.55.4067}{{\em Phys. Rev. E}
  {\bfseries 55} (1997)4067--4083}.

\bibitem{RevModPhys.53.385}
T.~A. Brody, J.~Flores, J.~B. French, P.~A. Mello, A.~Pandey, and S.~S.~M.
  Wong, ``Random-matrix physics: spectrum and strength fluctuations'',
  \href{http://dx.doi.org/10.1103/RevModPhys.53.385}{{\em Rev. Mod. Phys.}
  {\bfseries 53} (1981)385--479}.

\bibitem{Guhr:1997ve}
T.~Guhr, A.~Muller-Groeling, and H.~A. Weidenmuller, ``{Random matrix theories
  in quantum physics: Common concepts}'',
  \href{http://dx.doi.org/10.1016/S0370-1573(97)00088-4}{{\em Phys. Rept.}
  {\bfseries 299} (1998)189--425},
  \href{http://arxiv.org/abs/cond-mat/9707301}{{\ttfamily
  arXiv:cond-mat/9707301}}.

\bibitem{B1}
O.~Bohigas, M.~J. Giannoni, and C.~Schmit, ``Characterization of chaotic
  quantum spectra and universality of level fluctuation laws'',
  \href{http://dx.doi.org/10.1103/PhysRevLett.52.1}{{\em Phys. Rev. Lett.}
  {\bfseries 52} (Jan, 1984)1--4}.

\bibitem{B2}
O.~Bohigas, M.~Giannoni, and C.~Schmit, ``Spectral properties of the laplacian
  and random matrix theories'',
  \href{http://dx.doi.org/https://doi.org/10.1051/jphyslet:0198400450210101500}{{\em
  Journal de Physique Lettres} {\bfseries 45} 21, (1984)1015--1022}.

\bibitem{cmp/1103904878}
L.~Bunimovich, ``{On the ergodic properties of nowhere dispersing billiards}'',
  \href{http://dx.doi.org/cmp/1103904878}{{\em Communications in Mathematical
  Physics} {\bfseries 65} 3, (1979)295--312}.

\bibitem{doi:10.1063/1.1665596}
M.~C. Gutzwiller, ``Periodic orbits and classical quantization conditions'',
  \href{http://dx.doi.org/10.1063/1.1665596}{{\em Journal of Mathematical
  Physics} {\bfseries 12} 3, (1971)343--358}.

\bibitem{10.2307/79349}
M.~V. Berry and M.~Tabor, ``Level clustering in the regular spectrum'',
  \href{http://dx.doi.org/10.1098/rspa.1977.0140}{{\em Proceedings of the Royal
  Society of London. Series A, Mathematical and Physical Sciences} {\bfseries
  356} 1686, (1977)375--394}.

\bibitem{PhysRevLett.69.2188}
J.~Bolte, G.~Steil, and F.~Steiner, ``Arithmetical chaos and violation of
  universality in energy level statistics'',
  \href{http://dx.doi.org/10.1103/PhysRevLett.69.2188}{{\em Phys. Rev. Lett.}
  {\bfseries 69} (Oct, 1992)2188--2191}.

\bibitem{whatischaos}
Z.~Rudnick, ``{What is quantum chaos?}'', {\em Notices of the AMS} {\bfseries
  55} 1, (2008)32--34.

\bibitem{zelditchchaos}
S.~Zelditch, ``{Mathematics of Quantum Chaos in 2019}'',
  \href{http://dx.doi.org/10.1090/noti1958}{{\em Notices of the AMS} {\bfseries
  66} 9, (2019)}.

\bibitem{Katok}
A.~Katok and J.-M. Strelcy, \href{http://dx.doi.org/10.1007/BFb0099031}{{\em
  Invariant Manifolds, Entropy and Billiards; Smooth Maps with Singularities}}.
\newblock Springer Berlin Heidelberg, 1986.

\bibitem{Hedlund}
G.~A. Hedlund, ``On the metrical transitivity of the geodesics on closed
  surfaces of constant negative curvature'',
  \href{http://dx.doi.org/10.2307/1968495}{{\em Annals of Mathematics}
  {\bfseries 35} 4, (1934)787--808}.

\bibitem{Hopf1}
E.~Hopf, ``{Ergodic theory and the geodesic flow on surfaces of constant
  negative curvature}'', \href{http://dx.doi.org/bams/1183533160}{{\em Bulletin
  of the American Mathematical Society} {\bfseries 77} 6, (1971)863 -- 877}.

\bibitem{Hopf2}
E.~Hopf, ``{Statistik der geodätischen Linien in Mannigfaltigkeiten negativer
  Krümmung}'', {\em Ber. Verh. Sächs. Akad. Wiss.} {\bfseries 91}
  (1939)261--304.

\bibitem{Asnov}
G.~A. Hedlund, ``Geodesic flows on closed riemann manifolds with negative
  curvature'', {\em Proceedings of the Steklov Institute of Mathematics} 90,
  (1967).

\bibitem{Berger}
M.~Berger, \href{http://dx.doi.org/10.1007/978-3-642-18245-7}{{\em A Panoramic
  View of Riemannian Geometry}}.
\newblock Springer Berlin Heidelberg, 2007.

\bibitem{10.2307/1970079}
E.~P. Wigner, ``Characteristic vectors of bordered matrices with infinite
  dimensions'', \href{http://dx.doi.org/https://doi.org/10.2307/1970079}{{\em
  Annals of Mathematics} {\bfseries 62} 3, (1955)548--564}.

\bibitem{PhysRev.120.1698}
N.~Rosenzweig and C.~E. Porter, ````{R}epulsion of energy levels'' in complex
  atomic spectra'', \href{http://dx.doi.org/10.1103/PhysRev.120.1698}{{\em
  Phys. Rev.} {\bfseries 120} (Dec, 1960)1698--1714}.

\bibitem{PhysRev.123.1293}
R.~E. Trees, ````{R}epulsion of energy levels'' in complex atomic spectra'',
  \href{http://dx.doi.org/10.1103/PhysRev.123.1293}{{\em Phys. Rev.} {\bfseries
  123} (Aug, 1961)1293--1300}.

\bibitem{Haq19821086}
R.~Haq, A.~Pandey, and O.~Bohigas, ``Fluctuation properties of nuclear energy
  levels: Do theory and experiment agree?'',
  \href{http://dx.doi.org/10.1103/PhysRevLett.48.1086}{{\em Physical Review
  Letters} {\bfseries 48} 16, (1982)1086--1089}.

\bibitem{Gutzwiller_1990}
M.~C. Gutzwiller, \href{http://dx.doi.org/10.1007/978-1-4612-0983-6}{{\em Chaos
  in Classical and Quantum Mechanics}}.
\newblock Springer New York, 1990.

\bibitem{BALAZS1986109}
N.~Balazs and A.~Voros, ``Chaos on the pseudosphere'',
  \href{http://dx.doi.org/https://doi.org/10.1016/0370-1573(86)90159-6}{{\em
  Physics Reports} {\bfseries 143} 3, (1986)109--240}.

\bibitem{PhysRevB.43.8641}
D.~J.~E. Callaway, ``Random matrices, fractional statistics, and the quantum
  hall effect'', \href{http://dx.doi.org/10.1103/PhysRevB.43.8641}{{\em Phys.
  Rev. B} {\bfseries 43} (Apr, 1991)8641--8643}.

\bibitem{Sachdev:1992fk}
S.~Sachdev and J.~Ye, ``{Gapless spin fluid ground state in a random, quantum
  Heisenberg magnet}'',
  \href{http://dx.doi.org/10.1103/PhysRevLett.70.3339}{{\em Phys. Rev. Lett.}
  {\bfseries 70} (1993)3339},
  \href{http://arxiv.org/abs/cond-mat/9212030}{{\ttfamily
  arXiv:cond-mat/9212030}}.

\bibitem{Kitaev1}
A.~Kitaev, ``Hidden correlations in the {H}awking radiation and thermal
  noise'', 1998.
\newblock \url{https://online.kitp.ucsb.edu/online/joint98/kitaev/}.

\bibitem{Kitaev2}
A.~Kitaev, ``A simple model of quantum holography {I}'', 2015.
\newblock \url{https://online.kitp.ucsb.edu/online/entangled15/kitaev/}.

\bibitem{Kitaev3}
A.~Kitaev, ``A simple model of quantum holography {II}'', 2015.
\newblock \url{https://online.kitp.ucsb.edu/online/entangled15/kitaev2/}.

\bibitem{Saad:2018bqo}
P.~Saad, S.~H. Shenker, and D.~Stanford, ``{A semiclassical ramp in SYK and in
  gravity}'', \href{http://arxiv.org/abs/1806.06840}{{\ttfamily
  arXiv:1806.06840 [hep-th]}}.

\bibitem{Garcia-Garcia:2016mno}
A.~M. Garc\'\i{}a-Garc\'\i{}a and J.~J.~M. Verbaarschot, ``{Spectral and
  thermodynamic properties of the Sachdev-Ye-Kitaev model}'',
  \href{http://dx.doi.org/10.1103/PhysRevD.94.126010}{{\em Phys. Rev. D}
  {\bfseries 94} 12, (2016)126010},
  \href{http://arxiv.org/abs/1610.03816}{{\ttfamily arXiv:1610.03816
  [hep-th]}}.

\bibitem{You_2017}
Y.-Z. You, A.~W.~W. Ludwig, and C.~Xu, ``Sachdev-{Y}e-{K}itaev model and
  thermalization on the boundary of many-body localized fermionic
  symmetry-protected topological states'',
  \href{http://dx.doi.org/10.1103/PhysRevB.95.115150}{{\em Phys. Rev. B}
  {\bfseries 95} (Mar, 2017)115150},
  \href{http://arxiv.org/abs/1602.06964}{{\ttfamily arXiv:1602.06964
  [cond-mat.str-el]}}.

\bibitem{Cotler:2016fpe}
J.~S. Cotler, G.~Gur-Ari, M.~Hanada, J.~Polchinski, P.~Saad, S.~H. Shenker,
  D.~Stanford, A.~Streicher, and M.~Tezuka, ``{Black Holes and Random
  Matrices}'', \href{http://dx.doi.org/10.1007/JHEP05(2017)118}{{\em JHEP}
  {\bfseries 05} (2017)118}, \href{http://arxiv.org/abs/1611.04650}{{\ttfamily
  arXiv:1611.04650 [hep-th]}}. [Erratum: JHEP 09, 002 (2018)].

\bibitem{Gharibyan:2018jrp}
H.~Gharibyan, M.~Hanada, S.~H. Shenker, and M.~Tezuka, ``{Onset of Random
  Matrix Behavior in Scrambling Systems}'',
  \href{http://dx.doi.org/10.1007/JHEP07(2018)124}{{\em JHEP} {\bfseries 07}
  (2018)124}, \href{http://arxiv.org/abs/1803.08050}{{\ttfamily
  arXiv:1803.08050 [hep-th]}}. [Erratum: JHEP 02, 197 (2019)].

\bibitem{Maldacena:2001kr}
J.~M. Maldacena, ``{Eternal black holes in anti-de Sitter}'',
  \href{http://dx.doi.org/10.1088/1126-6708/2003/04/021}{{\em JHEP} {\bfseries
  04} (2003)021}, \href{http://arxiv.org/abs/hep-th/0106112}{{\ttfamily
  arXiv:hep-th/0106112}}.

\bibitem{Dyson:2002nt}
L.~Dyson, J.~Lindesay, and L.~Susskind, ``{Is there really a de Sitter/CFT
  duality?}'', \href{http://dx.doi.org/10.1088/1126-6708/2002/08/045}{{\em
  JHEP} {\bfseries 08} (2002)045},
  \href{http://arxiv.org/abs/hep-th/0202163}{{\ttfamily arXiv:hep-th/0202163}}.

\bibitem{Barbon:2003aq}
J.~L.~F. Barbon and E.~Rabinovici, ``{Very long time scales and black hole
  thermal equilibrium}'',
  \href{http://dx.doi.org/10.1088/1126-6708/2003/11/047}{{\em JHEP} {\bfseries
  11} (2003)047}, \href{http://arxiv.org/abs/hep-th/0308063}{{\ttfamily
  arXiv:hep-th/0308063}}.

\bibitem{Papadodimas:2015xma}
K.~Papadodimas and S.~Raju, ``{Local Operators in the Eternal Black Hole}'',
  \href{http://dx.doi.org/10.1103/PhysRevLett.115.211601}{{\em Phys. Rev.
  Lett.} {\bfseries 115} 21, (2015)211601},
  \href{http://arxiv.org/abs/1502.06692}{{\ttfamily arXiv:1502.06692
  [hep-th]}}.

\bibitem{Afkhami-Jeddi:2020ezh}
N.~Afkhami-Jeddi, H.~Cohn, T.~Hartman, and A.~Tajdini, ``{Free partition
  functions and an averaged holographic duality}'',
  \href{http://dx.doi.org/10.1007/JHEP01(2021)130}{{\em JHEP} {\bfseries 01}
  (2021)130}, \href{http://arxiv.org/abs/2006.04839}{{\ttfamily
  arXiv:2006.04839 [hep-th]}}.

\bibitem{Maloney:2020nni}
A.~Maloney and E.~Witten, ``{Averaging over Narain moduli space}'',
  \href{http://dx.doi.org/10.1007/JHEP10(2020)187}{{\em JHEP} {\bfseries 10}
  (2020)187}, \href{http://arxiv.org/abs/2006.04855}{{\ttfamily
  arXiv:2006.04855 [hep-th]}}.

\bibitem{Benjamin:2021wzr}
N.~Benjamin, C.~A. Keller, H.~Ooguri, and I.~G. Zadeh, ``{Narain to Narnia}'',
  \href{http://arxiv.org/abs/2103.15826}{{\ttfamily arXiv:2103.15826
  [hep-th]}}.

\bibitem{Witten:1982df}
E.~Witten, ``{Constraints on Supersymmetry Breaking}'',
  \href{http://dx.doi.org/10.1016/0550-3213(82)90071-2}{{\em Nucl. Phys. B}
  {\bfseries 202} (1982)253}.

\bibitem{Eguchi:1987wf}
T.~Eguchi and A.~Taormina, ``{Character Formulas for the $N=4$ Superconformal
  Algebra}'', \href{http://dx.doi.org/10.1016/0370-2693(88)90778-2}{{\em Phys.
  Lett. B} {\bfseries 200} (1988)315}.

\bibitem{Eguchi:1988vra}
T.~Eguchi, H.~Ooguri, A.~Taormina, and S.-K. Yang, ``{Superconformal Algebras
  and String Compactification on Manifolds with SU(N) Holonomy}'',
  \href{http://dx.doi.org/10.1016/0550-3213(89)90454-9}{{\em Nucl. Phys. B}
  {\bfseries 315} (1989)193--221}.

\bibitem{Cecotti:1992qh}
S.~Cecotti, P.~Fendley, K.~A. Intriligator, and C.~Vafa, ``{A New
  supersymmetric index}'',
  \href{http://dx.doi.org/10.1016/0550-3213(92)90572-S}{{\em Nucl. Phys. B}
  {\bfseries 386} (1992)405--452},
  \href{http://arxiv.org/abs/hep-th/9204102}{{\ttfamily arXiv:hep-th/9204102}}.

\bibitem{Kawai:1993jk}
T.~Kawai, Y.~Yamada, and S.-K. Yang, ``{Elliptic genera and N=2 superconformal
  field theory}'', \href{http://dx.doi.org/10.1016/0550-3213(94)90428-6}{{\em
  Nucl. Phys. B} {\bfseries 414} (1994)191--212},
  \href{http://arxiv.org/abs/hep-th/9306096}{{\ttfamily arXiv:hep-th/9306096}}.

\bibitem{Dijkgraaf:1996it}
R.~Dijkgraaf, E.~P. Verlinde, and H.~L. Verlinde, ``{Counting dyons in N=4
  string theory}'', \href{http://dx.doi.org/10.1016/S0550-3213(96)00640-2}{{\em
  Nucl. Phys. B} {\bfseries 484} (1997)543--561},
  \href{http://arxiv.org/abs/hep-th/9607026}{{\ttfamily arXiv:hep-th/9607026}}.

\bibitem{Benini:2013nda}
F.~Benini, R.~Eager, K.~Hori, and Y.~Tachikawa, ``{Elliptic genera of
  two-dimensional N=2 gauge theories with rank-one gauge groups}'',
  \href{http://dx.doi.org/10.1007/s11005-013-0673-y}{{\em Lett. Math. Phys.}
  {\bfseries 104} (2014)465--493},
  \href{http://arxiv.org/abs/1305.0533}{{\ttfamily arXiv:1305.0533 [hep-th]}}.

\bibitem{Lin:2015dsa}
Y.-H. Lin, S.-H. Shao, Y.~Wang, and X.~Yin, ``{Supersymmetry Constraints and
  String Theory on K3}'', \href{http://dx.doi.org/10.1007/JHEP12(2015)142}{{\em
  JHEP} {\bfseries 12} (2015)142},
  \href{http://arxiv.org/abs/1508.07305}{{\ttfamily arXiv:1508.07305
  [hep-th]}}.

\bibitem{Hellerman:2009bu}
S.~Hellerman, ``{A Universal Inequality for CFT and Quantum Gravity}'',
  \href{http://dx.doi.org/10.1007/JHEP08(2011)130}{{\em JHEP} {\bfseries 08}
  (2011)130}, \href{http://arxiv.org/abs/0902.2790}{{\ttfamily arXiv:0902.2790
  [hep-th]}}.

\bibitem{Keller:2012mr}
C.~A. Keller and H.~Ooguri, ``{Modular Constraints on Calabi-Yau
  Compactifications}'', \href{http://dx.doi.org/10.1007/s00220-013-1797-8}{{\em
  Commun. Math. Phys.} {\bfseries 324} (2013)107--127},
  \href{http://arxiv.org/abs/1209.4649}{{\ttfamily arXiv:1209.4649 [hep-th]}}.

\bibitem{Friedan:2013cba}
D.~Friedan and C.~A. Keller, ``{Constraints on 2d CFT partition functions}'',
  \href{http://dx.doi.org/10.1007/JHEP10(2013)180}{{\em JHEP} {\bfseries 10}
  (2013)180}, \href{http://arxiv.org/abs/1307.6562}{{\ttfamily arXiv:1307.6562
  [hep-th]}}.

\bibitem{Lin:2015wcg}
Y.-H. Lin, S.-H. Shao, D.~Simmons-Duffin, Y.~Wang, and X.~Yin, ``{$ \mathcal{N}
  $ = 4 superconformal bootstrap of the K3 CFT}'',
  \href{http://dx.doi.org/10.1007/JHEP05(2017)126}{{\em JHEP} {\bfseries 05}
  (2017)126}, \href{http://arxiv.org/abs/1511.04065}{{\ttfamily
  arXiv:1511.04065 [hep-th]}}.

\bibitem{Collier:2016cls}
S.~Collier, Y.-H. Lin, and X.~Yin, ``{Modular Bootstrap Revisited}'',
  \href{http://dx.doi.org/10.1007/JHEP09(2018)061}{{\em JHEP} {\bfseries 09}
  (2018)061}, \href{http://arxiv.org/abs/1608.06241}{{\ttfamily
  arXiv:1608.06241 [hep-th]}}.

\bibitem{Lin:2016gcl}
Y.-H. Lin, S.-H. Shao, Y.~Wang, and X.~Yin, ``{(2, 2) superconformal bootstrap
  in two dimensions}'', \href{http://dx.doi.org/10.1007/JHEP05(2017)112}{{\em
  JHEP} {\bfseries 05} (2017)112},
  \href{http://arxiv.org/abs/1610.05371}{{\ttfamily arXiv:1610.05371
  [hep-th]}}.

\bibitem{1810.10540}
S.~Kachru, A.~Tripathy, and M.~Zimet, ``{K3 metrics from little string
  theory}'', \href{http://arxiv.org/abs/1810.10540}{{\ttfamily arXiv:1810.10540
  [hep-th]}}.

\bibitem{2006.02435}
S.~Kachru, A.~Tripathy, and M.~Zimet, ``{K3 metrics}'',
  \href{http://arxiv.org/abs/2006.02435}{{\ttfamily arXiv:2006.02435
  [hep-th]}}.

\bibitem{2010.12581}
A.~Tripathy and M.~Zimet, ``{A plethora of K3 metrics}'',
  \href{http://arxiv.org/abs/2010.12581}{{\ttfamily arXiv:2010.12581
  [hep-th]}}.

\bibitem{hep-th/0506129}
M.~Headrick and T.~Wiseman, ``{Numerical Ricci-flat metrics on K3}'',
  \href{http://dx.doi.org/10.1088/0264-9381/22/23/002}{{\em Class. Quant.
  Grav.} {\bfseries 22} (2005)4931--4960},
  \href{http://arxiv.org/abs/hep-th/0506129}{{\ttfamily arXiv:hep-th/0506129}}.

\bibitem{hep-th/0612075}
M.~R. Douglas, R.~L. Karp, S.~Lukic, and R.~Reinbacher, ``{Numerical Calabi-Yau
  metrics}'', \href{http://dx.doi.org/10.1063/1.2888403}{{\em J. Math. Phys.}
  {\bfseries 49} (2008)032302},
  \href{http://arxiv.org/abs/hep-th/0612075}{{\ttfamily arXiv:hep-th/0612075}}.

\bibitem{0712.3563}
V.~Braun, T.~Brelidze, M.~R. Douglas, and B.~A. Ovrut, ``{Calabi-Yau Metrics
  for Quotients and Complete Intersections}'',
  \href{http://dx.doi.org/10.1088/1126-6708/2008/05/080}{{\em JHEP} {\bfseries
  05} (2008)080}, \href{http://arxiv.org/abs/0712.3563}{{\ttfamily
  arXiv:0712.3563 [hep-th]}}.

\bibitem{math/0512625}
S.~Donaldson, ``Some numerical results in complex differential geometry'',
  \href{http://dx.doi.org/10.4310/PAMQ.2009.v5.n2.a2}{{\em Pure and Applied
  Mathematics Quarterly} {\bfseries 5} 2, (2009)571--618},
  \href{http://arxiv.org/abs/math/0512625}{{\ttfamily arXiv:math/0512625
  [math.DG]}}.

\bibitem{0908.2635}
M.~Headrick and A.~Nassar, ``{Energy functionals for Calabi-Yau metrics}'',
  \href{http://dx.doi.org/10.4310/ATMP.2013.v17.n5.a1}{{\em Adv. Theor. Math.
  Phys.} {\bfseries 17} 5, (2013)867--902},
  \href{http://arxiv.org/abs/0908.2635}{{\ttfamily arXiv:0908.2635 [hep-th]}}.

\bibitem{2012.04656}
L.~B. Anderson, M.~Gerdes, J.~Gray, S.~Krippendorf, N.~Raghuram, and F.~Ruehle,
  ``{Moduli-dependent Calabi-Yau and SU(3)-structure metrics from Machine
  Learning}'', \href{http://dx.doi.org/10.1007/JHEP05(2021)013}{{\em JHEP}
  {\bfseries 05} (2021)013}, \href{http://arxiv.org/abs/2012.04656}{{\ttfamily
  arXiv:2012.04656 [hep-th]}}.

\bibitem{2012.04797}
M.~R. Douglas, S.~Lakshminarasimhan, and Y.~Qi, ``{Numerical Calabi-Yau metrics
  from holomorphic networks}'',
  \href{http://arxiv.org/abs/2012.04797}{{\ttfamily arXiv:2012.04797
  [hep-th]}}.

\bibitem{2012.15821}
V.~Jejjala, D.~K. Mayorga~Pena, and C.~Mishra, ``{Neural Network Approximations
  for Calabi-Yau Metrics}'', \href{http://arxiv.org/abs/2012.15821}{{\ttfamily
  arXiv:2012.15821 [hep-th]}}.

\bibitem{0805.3689}
V.~Braun, T.~Brelidze, M.~R. Douglas, and B.~A. Ovrut, ``{Eigenvalues and
  Eigenfunctions of the Scalar Laplace Operator on Calabi-Yau Manifolds}'',
  \href{http://dx.doi.org/10.1088/1126-6708/2008/07/120}{{\em JHEP} {\bfseries
  07} (2008)120}, \href{http://arxiv.org/abs/0805.3689}{{\ttfamily
  arXiv:0805.3689 [hep-th]}}.

\bibitem{Ashmore:2020ujw}
A.~Ashmore, ``{Eigenvalues and eigenforms on Calabi-Yau threefolds}'',
  \href{http://arxiv.org/abs/2011.13929}{{\ttfamily arXiv:2011.13929
  [hep-th]}}.

\bibitem{10.1007/978-3-0348-8266-8_36}
J.~Marklof, \href{http://dx.doi.org/10.1007/978-3-0348-8266-8_36}{``The
  {B}erry--{T}abor conjecture'',} in {\em European Congress of Mathematics},
  pp.~421--427.
\newblock Birkh{\"a}user Basel, 2001.

\bibitem{Hinterbichler:2013kwa}
K.~Hinterbichler, J.~Levin, and C.~Zukowski, ``{Kaluza-Klein Towers on General
  Manifolds}'', \href{http://dx.doi.org/10.1103/PhysRevD.89.086007}{{\em Phys.
  Rev. D} {\bfseries 89} 8, (2014)086007},
  \href{http://arxiv.org/abs/1310.6353}{{\ttfamily arXiv:1310.6353 [hep-th]}}.

\bibitem{Ashok:2003gk}
S.~Ashok and M.~R. Douglas, ``{Counting flux vacua}'',
  \href{http://dx.doi.org/10.1088/1126-6708/2004/01/060}{{\em JHEP} {\bfseries
  01} (2004)060}, \href{http://arxiv.org/abs/hep-th/0307049}{{\ttfamily
  arXiv:hep-th/0307049}}.

\bibitem{Douglas:2004zu}
M.~R. Douglas, B.~Shiffman, and S.~Zelditch, ``{Critical points and
  supersymmetric vacua}'',
  \href{http://dx.doi.org/10.1007/s00220-004-1228-y}{{\em Commun. Math. Phys.}
  {\bfseries 252} (2004)325--358},
  \href{http://arxiv.org/abs/math/0402326}{{\ttfamily arXiv:math/0402326}}.

\bibitem{Douglas:2003um}
M.~R. Douglas, ``{The Statistics of string / M theory vacua}'',
  \href{http://dx.doi.org/10.1088/1126-6708/2003/05/046}{{\em JHEP} {\bfseries
  05} (2003)046}, \href{http://arxiv.org/abs/hep-th/0303194}{{\ttfamily
  arXiv:hep-th/0303194}}.

\bibitem{Denef:2004ze}
F.~Denef and M.~R. Douglas, ``{Distributions of flux vacua}'',
  \href{http://dx.doi.org/10.1088/1126-6708/2004/05/072}{{\em JHEP} {\bfseries
  05} (2004)072}, \href{http://arxiv.org/abs/hep-th/0404116}{{\ttfamily
  arXiv:hep-th/0404116}}.

\bibitem{Denef:2004cf}
F.~Denef and M.~R. Douglas, ``{Distributions of nonsupersymmetric flux
  vacua}'', \href{http://dx.doi.org/10.1088/1126-6708/2005/03/061}{{\em JHEP}
  {\bfseries 03} (2005)061},
  \href{http://arxiv.org/abs/hep-th/0411183}{{\ttfamily arXiv:hep-th/0411183}}.

\bibitem{Distler:2005hi}
J.~Distler and U.~Varadarajan, ``{Random polynomials and the friendly
  landscape}'', \href{http://arxiv.org/abs/hep-th/0507090}{{\ttfamily
  arXiv:hep-th/0507090}}.

\bibitem{Podolsky:2008du}
D.~I. Podolsky, J.~Majumder, and N.~Jokela, ``{Disorder on the landscape}'',
  \href{http://dx.doi.org/10.1088/1475-7516/2008/05/024}{{\em JCAP} {\bfseries
  05} (2008)024}, \href{http://arxiv.org/abs/0804.2263}{{\ttfamily
  arXiv:0804.2263 [hep-th]}}.

\bibitem{Marsh:2011aa}
D.~Marsh, L.~McAllister, and T.~Wrase, ``{The Wasteland of Random
  Supergravities}'', \href{http://dx.doi.org/10.1007/JHEP03(2012)102}{{\em
  JHEP} {\bfseries 03} (2012)102},
  \href{http://arxiv.org/abs/1112.3034}{{\ttfamily arXiv:1112.3034 [hep-th]}}.

\bibitem{Eckerle:2016hzt}
K.~Eckerle and B.~Greene, ``{Random Field Theories in The Mirror Quintic Moduli
  Space}'', \href{http://arxiv.org/abs/1608.05189}{{\ttfamily arXiv:1608.05189
  [hep-th]}}.

\bibitem{Collier:2019weq}
S.~Collier, A.~Maloney, H.~Maxfield, and I.~Tsiares, ``{Universal dynamics of
  heavy operators in CFT$_{2}$}'',
  \href{http://dx.doi.org/10.1007/JHEP07(2020)074}{{\em JHEP} {\bfseries 07}
  (2020)074}, \href{http://arxiv.org/abs/1912.00222}{{\ttfamily
  arXiv:1912.00222 [hep-th]}}.

\bibitem{Cardy:1986ie}
J.~L. Cardy, ``{Operator Content of Two-Dimensional Conformally Invariant
  Theories}'', \href{http://dx.doi.org/10.1016/0550-3213(86)90552-3}{{\em Nucl.
  Phys. B} {\bfseries 270} (1986)186--204}.

\bibitem{1611.04592}
E.~Dyer and G.~Gur-Ari, ``{2D CFT Partition Functions at Late Times}'',
  \href{http://dx.doi.org/10.1007/JHEP08(2017)075}{{\em JHEP} {\bfseries 08}
  (2017)075}, \href{http://arxiv.org/abs/1611.04592}{{\ttfamily
  arXiv:1611.04592 [hep-th]}}.

\bibitem{1706.05400}
J.~Cotler, N.~Hunter-Jones, J.~Liu, and B.~Yoshida, ``{Chaos, Complexity, and
  Random Matrices}'', \href{http://dx.doi.org/10.1007/JHEP11(2017)048}{{\em
  JHEP} {\bfseries 11} (2017)048},
  \href{http://arxiv.org/abs/1706.05400}{{\ttfamily arXiv:1706.05400
  [hep-th]}}.

\bibitem{Cotler:2020ugk}
J.~Cotler and K.~Jensen, ``{AdS$_{3}$ gravity and random CFT}'',
  \href{http://dx.doi.org/10.1007/JHEP04(2021)033}{{\em JHEP} {\bfseries 04}
  (2021)033}, \href{http://arxiv.org/abs/2006.08648}{{\ttfamily
  arXiv:2006.08648 [hep-th]}}.

\bibitem{Saad:2019lba}
P.~Saad, S.~H. Shenker, and D.~Stanford, ``{JT gravity as a matrix integral}'',
  \href{http://arxiv.org/abs/1903.11115}{{\ttfamily arXiv:1903.11115
  [hep-th]}}.

\bibitem{Collier:2021rsn}
S.~Collier and A.~Maloney, ``{Wormholes and Spectral Statistics in the Narain
  Ensemble}'', \href{http://arxiv.org/abs/2106.12760}{{\ttfamily
  arXiv:2106.12760 [hep-th]}}.

\bibitem{Zirnbauer10}
M.~R. {Zirnbauer}, ``{Symmetry Classes}'',
  \href{http://arxiv.org/abs/1001.0722}{{\ttfamily arXiv:1001.0722 [math-ph]}}.

\bibitem{chao-dyn/9606010}
R.~Prange, ``The spectral form factor is not self-averaging'',
  \href{http://dx.doi.org/10.1103/PhysRevLett.78.2280}{{\em Physical review
  letters} {\bfseries 78} 12, (1997)2280},
  \href{http://arxiv.org/abs/chao-dyn/9606010}{{\ttfamily
  arXiv:chao-dyn/9606010}}.

\bibitem{Haake10}
F.~Haake, \href{http://dx.doi.org/10.1007/978-3-642-05428-0}{{\em Quantum
  Signatures of Chaos}}.
\newblock Springer Series in Synergetics. Springer, Berlin, Heidelberg, 3~ed.,
  2010.

\bibitem{Friedan:1980jf}
D.~Friedan, ``{Nonlinear Models in Two Epsilon Dimensions}'',
  \href{http://dx.doi.org/10.1103/PhysRevLett.45.1057}{{\em Phys. Rev. Lett.}
  {\bfseries 45} (1980)1057}.

\bibitem{AlvarezGaume:1981hm}
L.~Alvarez-Gaume and D.~Z. Freedman, ``{Geometrical Structure and Ultraviolet
  Finiteness in the Supersymmetric Sigma Model}'',
  \href{http://dx.doi.org/10.1007/BF01208280}{{\em Commun. Math. Phys.}
  {\bfseries 80} (1981)443}.

\bibitem{AlvarezGaume:1980dk}
L.~Alvarez-Gaume and D.~Z. Freedman, ``{Kahler Geometry and the Renormalization
  of Supersymmetric Sigma Models}'',
  \href{http://dx.doi.org/10.1103/PhysRevD.22.846}{{\em Phys. Rev. D}
  {\bfseries 22} (1980)846}.

\bibitem{AlvarezGaume:1981hn}
L.~Alvarez-Gaume, D.~Z. Freedman, and S.~Mukhi, ``{The Background Field Method
  and the Ultraviolet Structure of the Supersymmetric Nonlinear Sigma Model}'',
  \href{http://dx.doi.org/10.1016/0003-4916(81)90006-3}{{\em Annals Phys.}
  {\bfseries 134} (1981)85}.

\bibitem{AlvarezGaume:1985ww}
L.~Alvarez-Gaume and P.~H. Ginsparg, ``{Finiteness of Ricci Flat Supersymmetric
  Nonlinear Sigma Models}'', \href{http://dx.doi.org/10.1007/BF01229382}{{\em
  Commun. Math. Phys.} {\bfseries 102} (1985)311}.

\bibitem{Gross:1986iv}
D.~J. Gross and E.~Witten, ``{Superstring Modifications of Einstein's
  Equations}'', \href{http://dx.doi.org/10.1016/0550-3213(86)90429-3}{{\em
  Nucl. Phys. B} {\bfseries 277} (1986)1}.

\bibitem{Hull:1985at}
C.~M. Hull, ``{ULTRAVIOLET FINITENESS OF SUPERSYMMETRIC NONLINEAR SIGMA
  MODELS}'', \href{http://dx.doi.org/10.1016/0550-3213(85)90317-7}{{\em Nucl.
  Phys. B} {\bfseries 260} (1985)182--202}.

\bibitem{AlvarezGaume:1985xfa}
L.~Alvarez-Gaume, S.~R. Coleman, and P.~H. Ginsparg, ``{Finiteness of Ricci
  Flat $N=2$ Supersymmetric $\sigma$ Models}'',
  \href{http://dx.doi.org/10.1007/BF01211757}{{\em Commun. Math. Phys.}
  {\bfseries 103} (1986)423}.

\bibitem{Nemeschansky:1986yx}
D.~Nemeschansky and A.~Sen, ``{Conformal Invariance of Supersymmetric $\sigma$
  Models on Calabi-yau Manifolds}'',
  \href{http://dx.doi.org/10.1016/0370-2693(86)91394-8}{{\em Phys. Lett. B}
  {\bfseries 178} (1986)365--369}.

\bibitem{Gao:2013mn}
P.~Gao and M.~R. Douglas, ``{Geodesics on Calabi-Yau manifolds and winding
  states in nonlinear sigma models}'',
  \href{http://dx.doi.org/10.3389/fphy.2013.00026}{{\em Frontiers in Physics}
  {\bfseries 1} (1, 2013)26}, \href{http://arxiv.org/abs/1301.1687}{{\ttfamily
  arXiv:1301.1687 [hep-th]}}.

\bibitem{Figueroa-OFarrill:1997djj}
J.~M. Figueroa-O'Farrill, C.~Kohl, and B.~J. Spence, ``{Supersymmetry and the
  cohomology of (hyper)Kahler manifolds}'',
  \href{http://dx.doi.org/10.1016/S0550-3213(97)00548-8}{{\em Nucl. Phys. B}
  {\bfseries 503} (1997)614--626},
  \href{http://arxiv.org/abs/hep-th/9705161}{{\ttfamily arXiv:hep-th/9705161}}.

\bibitem{Singleton:2016hky}
A.~J. Singleton, \href{http://dx.doi.org/10.17863/CAM.275}{{\em {The geometry
  and representation theory of superconformal quantum mechanics}}}.
\newblock PhD thesis, Cambridge U., 6, 2016.

\bibitem{Weyl11}
H.~Weyl, ``{\"Uber die Asymptotische Verteilung der Eigenwertel}'', {\em Nachr.
  Konigl. Ges. Wiss.} (1911)110--117. \url{http://eudml.org/doc/58792}.

\bibitem{EigenvaluesinRiemannianGeometry}
I.~Chavel, \href{http://dx.doi.org/10.1016/S0079-8169(13)62888-3}{{\em
  Eigenvalues in Riemannian Geometry}}, vol.~115 of {\em Pure and Applied
  Mathematics}.
\newblock Elsevier, 1984.

\bibitem{Canzani13}
Y.~Canzani, ``Analysis on manifolds via the {L}aplacian'', 2013.
\newblock \url{http://www.math.harvard.edu/canzani/docs/Laplacian.pdf}.

\bibitem{1910.04767}
J.~Bonifacio and K.~Hinterbichler, ``{Unitarization from Geometry}'',
  \href{http://dx.doi.org/10.1007/JHEP12(2019)165}{{\em JHEP} {\bfseries 12}
  (2019)165}, \href{http://arxiv.org/abs/1910.04767}{{\ttfamily
  arXiv:1910.04767 [hep-th]}}.

\bibitem{2007.10337}
J.~Bonifacio and K.~Hinterbichler, ``{Bootstrap Bounds on Closed Einstein
  Manifolds}'', \href{http://dx.doi.org/10.1007/JHEP10(2020)069}{{\em JHEP}
  {\bfseries 10} (2020)069}, \href{http://arxiv.org/abs/2007.10337}{{\ttfamily
  arXiv:2007.10337 [hep-th]}}.

\bibitem{Bonifacio:2021msa}
J.~Bonifacio, ``{Bootstrap Bounds on Closed Hyperbolic Manifolds}'',
  \href{http://arxiv.org/abs/2107.09674}{{\ttfamily arXiv:2107.09674
  [hep-th]}}.

\bibitem{EinsteinManifolds}
A.~Besse, \href{http://dx.doi.org/10.1007/978-3-540-74311-8}{{\em Einstein
  Manifolds}}, vol.~115 of {\em Ergebnisse der Mathematik und ihrer
  Grenzgebiete}.
\newblock Springer, 1987.

\bibitem{1910.08605}
A.~Ashmore, Y.-H. He, and B.~A. Ovrut, ``{Machine Learning
  Calabi\textendash{}Yau Metrics}'',
  \href{http://dx.doi.org/10.1002/prop.202000068}{{\em Fortsch. Phys.}
  {\bfseries 68} 9, (2020)2000068},
  \href{http://arxiv.org/abs/1910.08605}{{\ttfamily arXiv:1910.08605
  [hep-th]}}.

\bibitem{Cui:2019uhy}
W.~Cui and J.~Gray, ``{Numerical Metrics, Curvature Expansions and Calabi-Yau
  Manifolds}'', \href{http://dx.doi.org/10.1007/JHEP05(2020)044}{{\em JHEP}
  {\bfseries 05} (2020)044}, \href{http://arxiv.org/abs/1912.11068}{{\ttfamily
  arXiv:1912.11068 [hep-th]}}.

\bibitem{0804.4555}
C.~Iuliu-Lazaroiu, D.~McNamee, and C.~Saemann, ``{Generalized Berezin
  quantization, Bergman metrics and fuzzy Laplacians}'',
  \href{http://dx.doi.org/10.1088/1126-6708/2008/09/059}{{\em JHEP} {\bfseries
  09} (2008)059}, \href{http://arxiv.org/abs/0804.4555}{{\ttfamily
  arXiv:0804.4555 [hep-th]}}.

\bibitem{2103.07472}
A.~Ashmore and F.~Ruehle, ``{Moduli-dependent KK towers and the swampland
  distance conjecture on the quintic Calabi-Yau manifold}'',
  \href{http://dx.doi.org/10.1103/PhysRevD.103.106028}{{\em Phys. Rev. D}
  {\bfseries 103} 10, (2021)106028},
  \href{http://arxiv.org/abs/2103.07472}{{\ttfamily arXiv:2103.07472
  [hep-th]}}.

\bibitem{Mathematica}
W.~R. Inc., ``Mathematica, {V}ersion 12.3.''
\newblock \url{https://www.wolfram.com/mathematica}. Champaign, IL, 2021.

\bibitem{Jockers:2012dk}
H.~Jockers, V.~Kumar, J.~M. Lapan, D.~R. Morrison, and M.~Romo, ``{Two-Sphere
  Partition Functions and Gromov-Witten Invariants}'',
  \href{http://dx.doi.org/10.1007/s00220-013-1874-z}{{\em Commun. Math. Phys.}
  {\bfseries 325} (2014)1139--1170},
  \href{http://arxiv.org/abs/1208.6244}{{\ttfamily arXiv:1208.6244 [hep-th]}}.

\bibitem{1206.2606}
N.~Doroud, J.~Gomis, B.~Le~Floch, and S.~Lee, ``{Exact Results in D=2
  Supersymmetric Gauge Theories}'',
  \href{http://dx.doi.org/10.1007/JHEP05(2013)093}{{\em JHEP} {\bfseries 05}
  (2013)093}, \href{http://arxiv.org/abs/1206.2606}{{\ttfamily arXiv:1206.2606
  [hep-th]}}.

\bibitem{1210.6022}
J.~Gomis and S.~Lee, ``{Exact Kahler Potential from Gauge Theory and Mirror
  Symmetry}'', \href{http://dx.doi.org/10.1007/JHEP04(2013)019}{{\em JHEP}
  {\bfseries 04} (2013)019}, \href{http://arxiv.org/abs/1210.6022}{{\ttfamily
  arXiv:1210.6022 [hep-th]}}.

\bibitem{Aleshkin:2017oak}
K.~Aleshkin and A.~Belavin, ``{A new approach for computing the geometry of the
  moduli spaces for a Calabi\textendash{}Yau manifold}'',
  \href{http://dx.doi.org/10.1088/1751-8121/aa9e7a}{{\em J. Phys. A} {\bfseries
  51} 5, (2018)055403}, \href{http://arxiv.org/abs/1706.05342}{{\ttfamily
  arXiv:1706.05342 [hep-th]}}.

\bibitem{Aleshkin:2017fuz}
K.~Aleshkin and A.~Belavin, ``{Special geometry on the 101 dimesional moduli
  space of the quintic threefold}'',
  \href{http://dx.doi.org/10.1007/JHEP03(2018)018}{{\em JHEP} {\bfseries 03}
  (2018)018}, \href{http://arxiv.org/abs/1710.11609}{{\ttfamily
  arXiv:1710.11609 [hep-th]}}.

\bibitem{Aleshkin:2018jql}
K.~Aleshkin and A.~Belavin, ``{Exact Computation of the Special Geometry for
  Calabi\textendash{}Yau Hypersurfaces of Fermat Type}'',
  \href{http://dx.doi.org/10.1134/S0021364018220010}{{\em JETP Lett.}
  {\bfseries 108} 10, (2018)705--709},
  \href{http://arxiv.org/abs/1806.02772}{{\ttfamily arXiv:1806.02772
  [hep-th]}}.

\bibitem{Keller:2009vj}
J.~Keller and S.~Lukic, ``{Numerical Weil\textendash{}Petersson metrics on
  moduli spaces of Calabi\textendash{}Yau manifolds}'',
  \href{http://dx.doi.org/10.1016/j.geomphys.2015.02.018}{{\em J. Geom. Phys.}
  {\bfseries 92} (2015)252--270},
  \href{http://arxiv.org/abs/0907.1387}{{\ttfamily arXiv:0907.1387 [math.DG]}}.

\bibitem{Candelas:1990rm}
P.~Candelas, X.~C. De~La~Ossa, P.~S. Green, and L.~Parkes, ``{A Pair of
  Calabi-Yau manifolds as an exactly soluble superconformal theory}'',
  \href{http://dx.doi.org/10.1016/0550-3213(91)90292-6}{{\em Nucl. Phys. B}
  {\bfseries 359} (1991)21--74}.

\bibitem{1803.04989}
R.~Blumenhagen, D.~Kl\"awer, L.~Schlechter, and F.~Wolf, ``{The Refined
  Swampland Distance Conjecture in Calabi-Yau Moduli Spaces}'',
  \href{http://dx.doi.org/10.1007/JHEP06(2018)052}{{\em JHEP} {\bfseries 06}
  (2018)052}, \href{http://arxiv.org/abs/1803.04989}{{\ttfamily
  arXiv:1803.04989 [hep-th]}}.

\bibitem{1905.05225}
D.~Erkinger and J.~Knapp, ``{Refined swampland distance conjecture and exotic
  hybrid Calabi-Yaus}'', \href{http://dx.doi.org/10.1007/JHEP07(2019)029}{{\em
  JHEP} {\bfseries 07} (2019)029},
  \href{http://arxiv.org/abs/1905.05225}{{\ttfamily arXiv:1905.05225
  [hep-th]}}.

\bibitem{1110.2150}
A.~{Strohmaier} and V.~{Uski}, ``{An Algorithm for the Computation of
  Eigenvalues, Spectral Zeta Functions and Zeta-Determinants on Hyperbolic
  Surfaces}'', \href{http://dx.doi.org/10.1007/s00220-012-1557-1}{{\em
  Communications in Mathematical Physics} {\bfseries 317} 3, (2013)827--869},
  \href{http://arxiv.org/abs/1110.2150}{{\ttfamily arXiv:1110.2150 [math.SP]}}.

\bibitem{Cook18}
J.~Cook, {\em
  \href{https://repository.lboro.ac.uk/articles/thesis/Properties_of_eigenvalues_on_Riemann_surfaces_with_large_symmetry_groups/9374180}{Properties
  of eigenvalues on Riemann surfaces with large symmetry groups}}.
\newblock PhD thesis, Loughborough University, 2018.

\bibitem{hep-th/9611137}
P.~S. Aspinwall, ``{K3 surfaces and string duality}'', in {\em {Theoretical
  Advanced Study Institute in Elementary Particle Physics (TASI 96): Fields,
  Strings, and Duality}}.
\newblock 11, 1996.
\newblock \href{http://arxiv.org/abs/hep-th/9611137}{{\ttfamily
  arXiv:hep-th/9611137}}.

\end{thebibliography}\endgroup

\end{document}